\documentclass[aps,prl,reprint,twocolumn,showpacs,floatfix,superscriptaddress]{revtex4-2}

\usepackage{amssymb,amsmath,amstext}                
\usepackage{graphicx}                                               

\usepackage{fp}
\newcount\boxheight
\newcount\boxwidth
\newcommand\testaspect[1]{%
  \setbox0=\hbox{#1}%
  \boxheight=\ht0\relax%
  \boxwidth=\wd0\relax%
  \FPdiv\theaspect{\the\boxheight}{\the\boxwidth}%
  \copy0%
}

\usepackage{epstopdf}                                               
\usepackage{color}                                                     
\usepackage{bm}                                                        
\usepackage{comment}     
\usepackage{appendix}                                              
\usepackage[utf8]{inputenc}
\usepackage{bbold}
\usepackage{bbm}
\usepackage{latexsym}
\usepackage[T1]{fontenc}
\usepackage{xcolor}
\usepackage{braket}
\definecolor{lblue} {RGB}{51,71,158}
\usepackage[colorlinks=true,citecolor=blue,linkcolor=blue,urlcolor=lblue]{hyperref}

\usepackage[normalem]{ulem}

\definecolor{darkgreen}{rgb}{0.13, 0.55, 0.13}

\def\beq{\begin{equation}}
\def\eeq{\end{equation}}

\usepackage{dcolumn}
\usepackage{sidecap}
\usepackage[caption=false]{subfig}
\usepackage{bm}
\usepackage{braket}
\begin{document}

\title{Phenomenology of many-body localization in bond-disordered spin chains}

\author{Adith Sai Aramthottil} 
\affiliation{Szkoła Doktorska Nauk \'Scis\l{}ych i Przyrodniczych, Uniwersytet Jagiello\'nski,  \L{}ojasiewicza 11, PL-30-348 Krak\'ow, Poland}
\affiliation{Instytut Fizyki Teoretycznej, 
Uniwersytet Jagiello\'nski,  \L{}ojasiewicza 11, PL-30-348 Krak\'ow, Poland}
\author{Piotr Sierant} 
\affiliation{ICFO-Institut de Ci\`encies Fot\`oniques, The Barcelona Institute of Science and Technology, Av. Carl Friedrich
Gauss 3, 08860 Castelldefels (Barcelona), Spain}
\author{Maciej Lewenstein} 
\affiliation{ICFO-Institut de Ci\`encies Fot\`oniques, The Barcelona Institute of Science and Technology, Av. Carl Friedrich
Gauss 3, 08860 Castelldefels (Barcelona), Spain}
\affiliation{ICREA, Passeig Lluis Companys 23, 08010 Barcelona, Spain}
\author{Jakub Zakrzewski} 
\affiliation{Instytut Fizyki Teoretycznej, 
Uniwersytet Jagiello\'nski,  \L{}ojasiewicza 11, PL-30-348 Krak\'ow, Poland}
\affiliation{Mark Kac Complex Systems Research Center, Uniwersytet Jagiello{\'n}ski, PL-30-348 Krak{\'o}w, Poland}

\begin{abstract}
Many-body localization (MBL) hinders the thermalization of quantum many-body systems in the presence of strong disorder. In this work, we study the MBL regime in bond-disordered spin-1/2 XXZ spin chain, finding the multimodal distribution of entanglement entropy in eigenstates, sub-Poissonian level statistics, and revealing a relation between operators and initial states required for examining the breakdown of thermalization in the time evolution of the system. We employ a real space renormalization group scheme to identify these phenomenological features of the MBL regime that extend beyond the standard picture of local integrals of motion relevant for systems with disorder coupled to on-site operators. Our results pave the way for experimental probing of MBL in bond-disordered spin chains. 
\end{abstract}
\maketitle

\paragraph*{Introduction.}
The Eigenstate thermalization hypothesis (ETH)~\cite{Deutsch91, Srednicki94, Srednicki99, Dalessio16} conjectures that quantum many-body systems thermalize and reach equilibrium state determined solely by a few macroscopic quantities, irrespective of the details of the initial state.  
Many-body localization (MBL)~\cite{Basko06, Gornyi05, Pal10, Nandkishore15, Alet18, Abanin19,Sierant24rev} has been proposed as a mechanism to avoid thermalization for a large class of strongly disordered systems, including spin \cite{Santos04, Oganesyan07, Luitz15}, bosonic \cite{Sierant17, Sierant18, Orell19, Hopjan20}, fermionic \cite{Mondaini15,Prelovsek16,Zakrzewski18,Kozarzewski18} and Floquet systems \cite{Lazarides15, Ponte15a, Ponte15b, Abanin16,Zhang16, Bairey17,Sahay21, Garratt21, Sonner21, Sierant23}.
\begin{figure}
\begin{center}
\includegraphics[width=\linewidth]{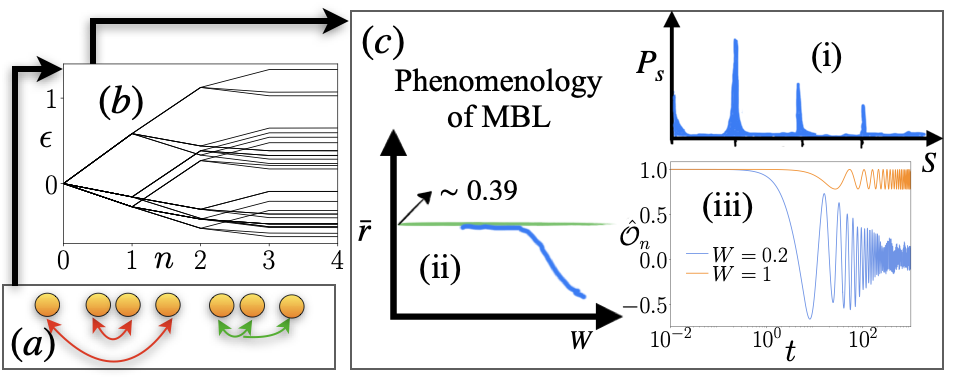}\\
\caption{
The RSRG-X scheme approximates the eigenstates and eigenvalues of the bond-disorder XXZ spin chain, capturing the phenomenology of the MBL regime at large disorder strength $W$.
(a) RSRG-X identifies and decimates the bonds between spins approximating eigenvalues ($\epsilon$) for a single realization of Hamiltonian \eqref{eq::pd_Ham} with $M=8$ spins, see panel (b). 
(c) Properties captured by the RSRG-X procedure: (i) nontrivial entanglement entropy distribution, (ii) sub-Poissonian mean gap ratio, (iii) lack of thermalization of $\hat{\mathcal O}_n$ at large $W$. 
} \label{fig:decimate_eg}
\end{center}
\end{figure}
Explorations of MBL have been primarily driven by numerical studies restricted by the computationally accessible system sizes~\cite{Pietracaprina18, Sierant20p} and timescales
~\cite{Bera17, Doggen18, Chanda20t}. Consequently, it is unclear whether the observed MBL \textit{regime} remains stable in the asymptotic limit of infinite system size and infinite time~\cite{Suntajs20,Sierant20b,Abanin21,Panda20}, motivating studies of persistent slow dynamics and prevalent system size drifts of ergodicity breaking indicators in strongly disordered interacting systems~\cite{Suntajs20e,  Kiefer20, Kiefer21, Sierant21constrained, Sels21a, Morningstar22, Sels22bath, Krajewski22, Sierant22}, as reviewed in \cite{Sierant24rev}.

The MBL regime is characterized, on experimentally relevant timescales and system sizes, by the slow growth of entanglement in dynamics \cite{Znidaric08, Bardarson12,Nanduri14,Lukin19}, power-law decay of local correlation functions \cite{Serbyn14,Schreiber15}, Poissonian level statistics~\cite{Serbyn16, Bertrand16, Sierant19b, Buijsman18, Sierant20}, and area-law entanglement of eigenstates \cite{Serbyn13a,Bauer13}. Local integrals of motions (LIOMs)~\cite{Serbyn13b,Huse14} provide a unifying framework that captures the phenomenology of MBL~\cite{Serbyn13a,Serbyn13b, Abanin19} in the vast class of systems~\cite{Santos04, Oganesyan07, Luitz15, Mondaini15, Prelovsek16, Sierant17, Sierant18,Orell19,Hopjan20, Zakrzewski18, Kozarzewski18, Lazarides15, Ponte15a, Ponte15b, Abanin16, Zhang16,Bairey17, Sahay21, Garratt21, Sonner21, Sierant23}. The LIOMs in the MBL regime can be constructed as long-time averages of local operators \cite{Chandran15}, with perturbative approaches \cite{Ros15}, with flow equation methods~\cite{Rademaker16,Monthus16, Thomson18, Thomson23} or direct numerical optimization~\cite{Brien16, Peng19, Johns19, Adami22}. In the paradigmatic disordered XXZ spin chain, LIOMs, $\hat \tau_i^z$, are the $z$-component of the spin operator $\hat s^z_i$ transformed by a quasilocal unitary~\cite{Ros15,Imbrie16,Abanin19}.
This paradigm applies whenever the disorder is coupled to on-site operators, which become exact LIOMs in the strong disorder limit.

This work focuses on many-body systems in which the disorder \textit{is not} coupled to on-site operators. We argue that the standard picture of LIOMs does not capture the phenomenology of MBL in such systems. Instead, the real space renormalization group for excited states (RSRG-X) ~\cite{Pekker14}, relevant for Hamiltonians constructed out of local operators with a broad distribution of energy gaps, provides a phenomenological description of the system's behavior at large disorder strengths. Instead, the real space renormalization group for excited states (RSRG-X)~\cite{Pekker14} provides a phenomenological description of the system's behavior at large disorder strengths. As a concrete example, we consider a random-bond XXZ spin-1/2 chain relevant for experiments with ultracold Rydberg atoms \cite{Whitlock19,Signoles21,Franz22,Franz22b}. For the considered model, the RSRG-X predicts sub-Poissonian level statistics, multimodal structure of the entanglement entropy distribution in eigenstates at large disorder strengths, and allows for identification of operators that may be conserved in the time evolution of experimentally relevant initial states and hence serve as order parameters for the MBL regime, see Fig.~\ref{fig:decimate_eg}. We show that the non-trivial memory effects of the RSRG-X procedure prevent us from finding a global set of LIOMs for the considered system even in the infinite disorder limit. We discuss how to extend our procedure to obtain experimentally verifiable predictions for other types of many-body systems.

\paragraph*{The model.}
We consider the random-bond XXZ spin chain 
\begin{equation}
    \hat{\mathcal{H}}= \sum_{i=1}^{M-1} J_{i,i+1}(\hat{s}_i^x\hat{s}_{i+1}^x + \hat{s}_i^y\hat{s}_{i+1}^y +\Delta \hat{s}_i^z\hat{s}_{i+1}^z) 
    \label{eq::pd_Ham}
\end{equation}
with open boundary conditions, where, $\hat{s}^{\beta}_i$ are the spin-$1/2$ operators acting on the $i$ site with $\beta \in \lbrace x,y,z \rbrace $, $J_{i,i+1}$ are couplings and $M$ is the number of spins.
We focus exclusively on the zero-magnetization sector and avoid the special case of the non-abelian SU(2) symmetry by imposing $\vert \Delta \vert  \neq 1$ \cite{Protopopov20}.
The remaining symmetry of the model is $\mathbb{Z}_2$ spin inversion symmetry $\mathcal{C}$. The Hamiltonian \eqref{eq::pd_Ham} is relevant for experiments with Rydberg atoms confined in an array of microtraps \cite{Whitlock17,Robert-de-Saint-Vincent13}. Investigations of these models have demonstrated a weak breaking of the ETH in the presence of interactions \cite{Franz22b} and glassy dynamics of selected operators \cite{Signoles21,Franz22}. With that motivation in mind, we assume that the couplings in \eqref{eq::pd_Ham} take values according to a power-law $J_{i,i+1}=\vert x_i -x_{i+1} \vert^{-\alpha}$, where $x_i$ denotes the position of the $i$-th spin within the length $L$ occupied by the system, and $\Delta =-0.73 $, $\alpha=6$ relevant for a Rydberg state of Rubidium 87. Fixing the Rydberg blockade~\cite{Lukin01} diameter as $D_{rb}$ results in 
an excluded density $\rho=D_{rb}M/L$.
With decreasing density $\rho$, spins can explore more positional configurations within $L$ there are more possible configurations present, leading to an increase of the effective disorder strength $W=1/\rho-1$, 
associated with the random distribution of $J_{i,i+1}$, in the model \cite{Braemer22} and thus to the MBL regime. 
In \cite{suppl_SD}, we describe our algorithm to generate particle configurations $\{ x_i \}$ avoiding the issue of the R{\'e}nyi parking constant \cite{Renyi58}.

\paragraph*{Decimation Scheme.}

To understand the MBL phenomenology of \eqref{eq::pd_Ham}, we employ the RSRG-X scheme~\cite{Pekker14}. 
The scheme proceeds iteratively by identifying the largest energy gap, diagonalizing the corresponding local operator, setting the state of the relevant spins into one of the obtained eigenstates, and perturbatively modifying the nearby effective local operators and the associated energy gaps~\cite{Dasgupta80,Pekker14}. 
The energy gaps encountered at each step of this procedure, which we refer to as a decimation, are progressively smaller. Hence, the chosen eigenstates fix the time dynamics at longer times, eventually producing an approximation of the eigenstate of \eqref{eq::pd_Ham} when the last spin is fixed. If the distribution of energy gaps is steadily broadening in the subsequent decimation steps, the scheme is increasingly well controlled and may become asymptotically exact~\cite{Fisher94, Vosk13, Turkeshi20negativity, Ruggiero22}. Here, relying on the broadness of the initial gap distribution, the RSRG-X is employed to obtain insights about relatively small systems relevant for numerical and experimental studies.

The RSRG-X proceeds as follows. The Hamiltonian \eqref{eq::pd_Ham} can be written as a sum of 2-site local operators of the form
\begin{equation}
    \hat{\mathcal{H}}_{i,i+1} = J_{i,i+1} \biggl(\frac{(\hat{s}_i^+\hat{s}_{i+1}^- +h.c )}{2}+\Delta_{i,i+1} \hat{s}_i^z\hat{s}_{i+1}^z\biggl),
    \label{eq::Ham_2s}
\end{equation}
with eigenbasis $ \vert Z_{\pm}\rangle  , \vert \pm \rangle$ where, $\vert Z_{+/-}\rangle \equiv \vert \uparrow  \uparrow  \rangle $/$\vert \downarrow  \downarrow  \rangle $,  and $ \vert \pm \rangle \equiv \frac{1}{\sqrt{2}}(\vert \uparrow \downarrow \rangle \pm \vert \downarrow \uparrow \rangle)$ and the parameter $\Delta_{i,i+1}=\Delta$. The relevant energy gap associated with each local operator is taken as the maximum of its energy gaps, so that any of the eigenstates can be chosen freely. This is a valid approximation if the energy scales are controlled by the couplings $J_{i,i+1}$. Let us assume that the largest relevant energy gap is associated with the local operator $\hat{\mathcal{H}}_{4,5}$. In that case, we select the bond between spins $i=3$, $i=4$ and fix the state of the two spins as $ \vert Z_{\pm}\rangle, \vert \pm \rangle$. We now describe how the neighboring operators, i.e., $\hat{\mathcal{H}}_{2,3}$ and $\hat{\mathcal{H}}_{4,5}$ are perturbatively modified for each of the choices.

For states $\vert \pm \rangle $, there are no contributions to the nearest operators within the first-order perturbation. The second-order perturbation results in the formation of a bond between spins $i=2$, $i=5$, described by $\hat{\mathcal{H}}_{2,5}$ of the form as in \eqref{eq::Ham_2s} with couplings modified to 
$\tilde{J}_{2,5} = \frac{J_{2,3}J_{4,5}}{(1\mp \Delta_{3,4})J_{3,4}} $, 
$\tilde{\Delta}_{2,5}= \frac{\mp\Delta_{2,3}\Delta_{4,5}(1\mp \Delta_{3,4})}{2}$. 
The choice $ \vert Z_{\pm}\rangle$, within the first-order perturbation modifies the operators $\hat{\mathcal{H}}_{2,3}$ and $\hat{\mathcal{H}}_{4,5}$ to take a chemical potential form \cite{Vasseur16} 
\begin{equation}
    \hat{\mathcal{H}}_{2,3}'=\pm\frac{1}{2}( \Delta_{2,3} J_{2,3}\hat{s}_{2}^z), \hat{\mathcal{H}}_{4,5}'= \pm\frac{1}{2}( \Delta_{4,5} J_{4,5}\hat{s}_{5}^z),
    \label{eq:1st_order}
\end{equation}
while the second order term leads to a bond between $i=2,5$ with $\tilde{J}_{2,5} =-\frac{J_{2,3}J_{4,5}}{2J_{3,4}}\left(\frac{1}{(1-\Delta_{3,4})}+\frac{1}{(1+\Delta_{3,4})}\right)$ and $\tilde{\Delta}_{2,5}=0$.
The first-order modification of $\hat{\mathcal{H}}'$ is typically much stronger than the second-order bond and the neighboring operators admit the chemical potential form in the following steps of the scheme. For instance, a decimation of the bond between sites $i=4$, $i=5$, would result in fixing the eigenstate at site $5$ to $\vert \uparrow \rangle /\vert \downarrow \rangle $ and modifying the next operator as $\hat{\mathcal{H}}_{5,6}'=\frac{1}{2}( \Delta_{5,6} J_{5,6}\hat{s}_{6}^z)/-\frac{1}{2}( \Delta_{5,6} J_{5,6}\hat{s}_{6}^z)$. 
This may lead to a process of growth of domains of spins fixed in states $\vert \uparrow \rangle /\vert \downarrow \rangle $ which stops if the next operator to be modified already has the form $\hat{\mathcal{H}}'$.
The spin polarized domains $\vert \uparrow \rangle /\vert \downarrow \rangle $ may extend over already decimates pair of sites with states fixed as $\vert \pm \rangle $.
 \begin{figure*}
\includegraphics[width=\linewidth]{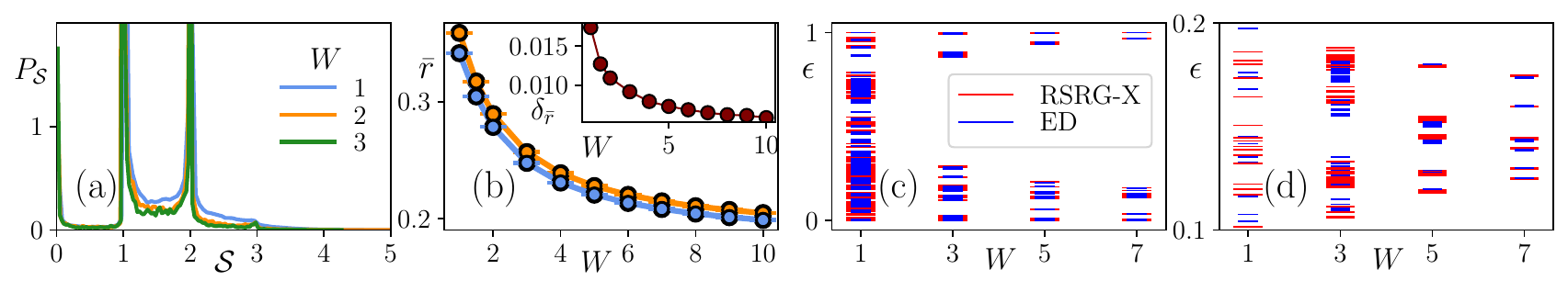}
\caption{
(a) Distribution ${P}_{\mathcal{S}}$ of half-chain entanglement entropy ${\mathcal{S}}$ for $M=20$ and varying disorder strength $W$.
(b) The mean gap ratio obtained by RSRG-X ($\bar{r}_{RG}$, blue) is compared with ED results ($\bar{r}_{ED}$, orange). The inset shows $\delta_{\bar{r}}=\bar{r}_{ED}-\bar{r}_{RG}$. The system size is $M=10$, with $25000$ disorder realizations and all the eigenvalues are considered. (c) The rescaled energies $\epsilon$ obtained with ED are compared with RSRG-X result for a single disorder realization, (d) is a zoomed image of the same. All the results for the  $\mathcal{C}=+1$ symmetry sector.
}\label{fig:work_dec} 
\end{figure*}
The generation of $\hat{\mathcal{H}'}$ operators through the first order perturbation suffices in all scenarios except when degenerate states $\vert K_+ \rangle \equiv \vert \uparrow \rangle\vert Z_- \rangle \vert \uparrow \rangle $, $\vert K_- \rangle \equiv \vert \downarrow \rangle\vert Z_+ \rangle \vert \downarrow \rangle $ are formed. Then, the second order perturbation needs to be taken into account leading to operators of the form $\hat{\mathcal{H}}$ with modified parameters, see~\cite{suppl_SD}.

A generic approximation of an eigenstate stemming from RSRG-X is associated with arbitrary choices of $\vert Z_{\pm} \rangle $, and $\vert\pm \rangle$ throughout the decimation scheme. To qualitatively understand the properties of these eigenstates, we start by assuming that only states $\vert \pm \rangle $ have been chosen in the decimation scheme. 
This leads to formation of bonds $\hat{\mathcal{H}}_{l_i,r_i}$ between arbitrarily distant spins, resulting in states $\vert \pm \rangle_{l_i,r_i}$. The spatial distribution of the correlations of states within this subsystem is analogous to a random singlet phase found for the ground state of \eqref{eq::pd_Ham}~\cite{Dasgupta80,Fisher94}, or for the excited states in the absence of interactions ($\Delta=0$)~\cite{Huang14,Pouranvari15}. 
In contrast,  fixing of the eigenstate as $\vert Z_{\pm} \rangle$ in the decimation procedure leads to growth of a domain which breaks the spin inversion symmetry $\vert \uparrow / \downarrow \rangle_{l_m}\cdots\vert \uparrow /\downarrow  \rangle_{l_2} \vert Z_{+/-} \rangle_{l_1,r_1}\vert \uparrow /\downarrow \rangle_{r_2}\cdots\vert \uparrow / \downarrow \rangle_{r_{m'}}$ (except for the configurations involving $\vert K_+ \rangle $ and $\vert K_- \rangle$). 
When the energy scales are not controlled by the couplings $J$, i.e., when  $\Delta \gg 1$, such spin inversion symmetry breaking domains extend through the entire system~\cite{Vasseur16}. In contrast, in our case, a generic eigenstate is formed by domains with spins polarized along the $z$ direction and subsystems in which the state is a product of triplet/singlet states $\vert \pm \rangle$ \footnote{ 
With probability $\approx 0.02$, the illustrated subsystem of states $\vert \pm \rangle$ contains states of the form $\frac{1}{\sqrt{2}}(\vert K_+\rangle \pm \vert K_-\rangle )$ }.

\paragraph*{Eigenstates and eigenvalues. }
To demonstrate that the eigenstates of the random-bond XXZ spin chain \eqref{eq::pd_Ham} at large $W$ indeed feature the structure predicted by the RSRG-X, we analyze the half-chain entanglement entropy $\mathcal{S}=-\mathrm{tr}(\rho \log_2(\rho) )$, where $\rho = \mathrm{tr_B}(\ket{\psi}\bra{\psi})$ is the reduced density obtained by tracing out the spins in half of the system and $\ket{\psi}$ is the considered eigenstate.
The distribution of $\mathcal{S}$, calculated by exact diagonalization (ED) and shown in Fig.~\ref{fig:work_dec}(a), exhibits peaks at integer values. The peaks correspond to different numbers of the bonds in triplet/singlet states $\ket{\pm}$ cut by the bipartition of the system, confirming the intuitive picture predicted by the RSRG-X. The peaks get broadened  at smaller disorder strengths $W$ due to the contribution of the higher order terms not taken into the account by our decimation scheme. 
The observed distribution of $\mathcal{S}$ is starkly different from the XXZ spin-1/2 chain which, in the large $W$ limit, collapses to a single peak at $\mathcal{S}=0$~\cite{Luitz16long}. 

The RSRG-X 
also approximates the eigenvalues $\epsilon_k$ of ~\eqref{eq::pd_Ham}. We compare the RSRG-X prediction with the ED results by calculating the gap ratio $r_k \equiv \frac{min\lbrace\delta_k,\delta_{k+1}\rbrace}{max\lbrace\delta_k,\delta_{k+1}\rbrace}$ (here $\delta_k \equiv \epsilon_{k+1} -\epsilon_{k}$) averaged over the entire spectrum of the system and disorder realizations, shown in Fig.~\ref{fig:work_dec}(b).
The RSRG-X qualitatively reproduces the behavior of $\bar{r}$, and, in particular, predicts the sub-Poissonian value $\bar{r} <0.386=\bar{r}_{PS}$ at larger $W$, where $ \bar{r}_{PS}\approx0.386$~\cite{Atas13} is the value for Poissonian level statistics found in the MBL regime of models with a complete set of LIOMs~\cite{Abanin19}. The sub-Poissonian level statistics arises due to multiple quasidegeneracies in the spectrum, qualitatively captured by the RSRG-X procedure. A closer inspection reveals that the $\bar{r}$ from the RSRG-X is systematically underestimating the ED value, which we attribute to higher order processes neglected by the RSRG-X. In~\cite{suppl_SD}, we show that this discrepancy is smaller when the distribution of $J$'s is wider, i.e., given by a power-law with $\alpha > 6$.

\paragraph*{Memory effects along the RSRG-X branch.}
Each step $n$ of the RSRG-X is associated with choice of the states $\vert Z_{\pm} \rangle $, $\vert \pm \rangle $ or, in case of the single-site decimations, with $ \ket{ \uparrow/\downarrow}$. While fixing $\ket{\pm}$ results only in decimation of the involved spins and renormalization of the parameters of the bond linking their neighbors, the selection of $\ket{Z_{\pm}}$ affects non-trivially the subsequent steps of the scheme. Each choice of $\ket{ Z_{\pm}}$ introduces a possibility of the growth of the domain of spins polarized along $z$ direction. This may arise whenever the interactions are present in the system, i.e., when $|\Delta_{i,i+1}|>0$. 
The triplet/singlet states $\ket{\pm}$ can not extend over the already fixed spin polarized domains.
Consequently, various branches of RSRG-X lead to eigenstates corresponding to different partitions of the system into spin polarized and triplet/singlet domains. Therefore, in contrast to the models in which the disorder is coupled to the on-site operators, we cannot find a complete set of (quasi-)local operators that would commute with the Hamiltonian \eqref{eq::pd_Ham}.

\begin{figure}
\begin{center}
\includegraphics[width=\linewidth]{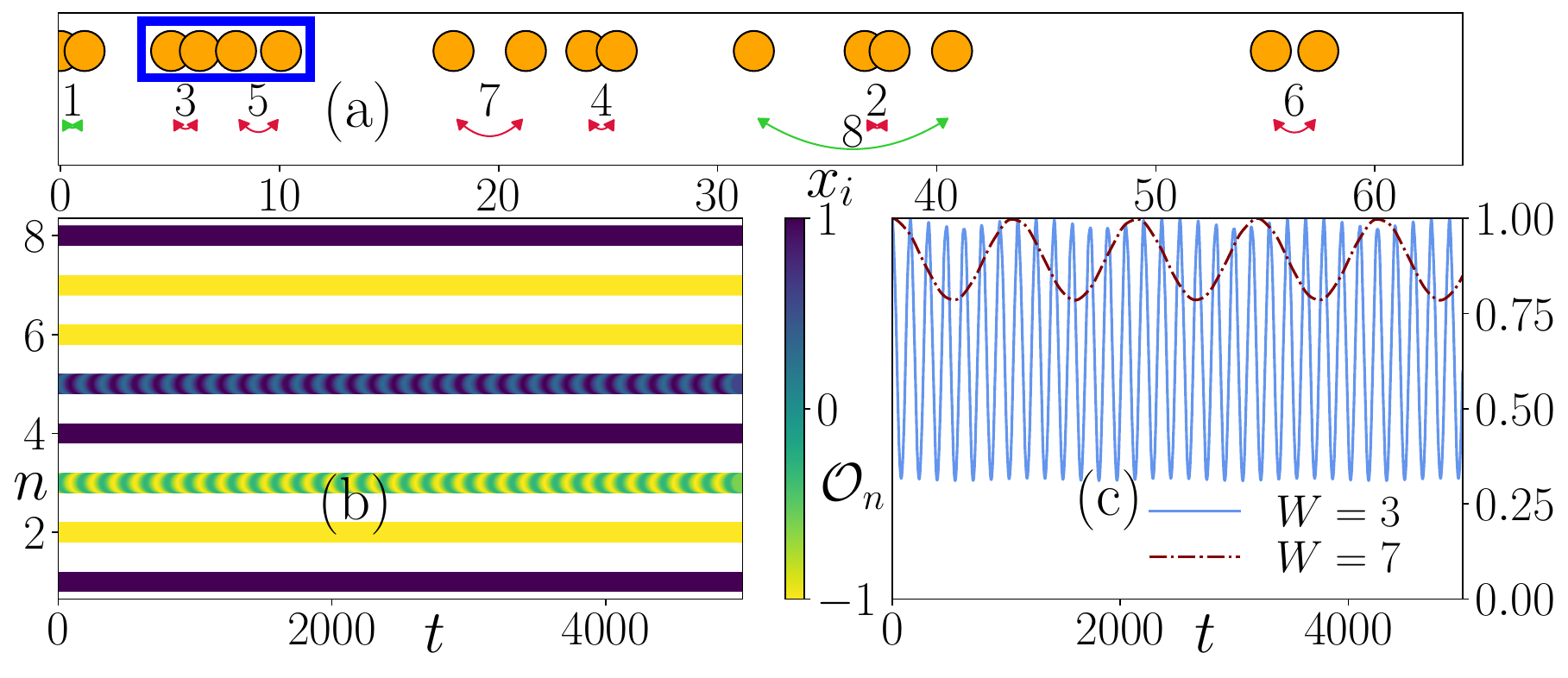}\\
 \caption{ 
(a) The decimation scheme for one configuration of spins at $W=3$ and $M=16$; connected spins represent decimation steps with $\ket{\pm}$ (in red) or $\ket{Z_{\pm}}$ (in green) fixed, and the numbers denote the number $n$ of the step; the blue box denotes an ergodic bubble (b) Time evolution of the average values $\mathcal O_n$ of operators 
corresponding to the RSRG-X branch shown in (a). (c) Time evolution of 
$\mathcal{O}_n(t)$ for the step $n=5$; the bold (dashed) line corresponds to $W=3$ ($W=7$).
}  \label{fig:sing_dis}
\end{center}
\end{figure}

\paragraph*{Probing the MBL regime with time dynamics. } 
Time evolution of the average value $s(t)$ of $\hat{s}^z_i$ operator is experimentally realizable way of studying the breakdown of thermalization in the standard disordered XXZ spin chain.
For an initial state taken as the eigenstate $\ket{\psi_Z}$ of $\hat{s}^z_i$ operators, $s(t)$ decays to zero in the ergodic regime, while, deep in the MBL regime, for large $W$, $s(t)$ remains close to unity even at long times \cite{Serbyn14,Schreiber15,Sierant22}. This procedure, however, would not work for the bond-disordered XXZ spin chain. Due to the non-trivial relations between decimations along various branches of RSRG-X and the presence of eigenstates at any energy density with triplet/singlet domains, the average $s(t)$ may dephase to zero irrespective of the disorder strength $W$.

To circumvent this issue, we propose to set as an initial state, $\ket{\psi_{\mathrm{in}}}$, the approximation of eigenstate obtained with RSRG-X and to measure average values of operators $\hat{\mathcal{O}}_n$  related to the RSRG-X branch leading to $\ket{\psi_{\mathrm{in}}}$. 
For decimations involving $\vert\pm \rangle$ we choose $\hat{\mathcal{O}}_n=\hat{T}_{i,j} \equiv (\hat{s}_{i}^+ \hat{s}_{j}^- +h.c.)$, while for $\vert Z_{\pm} \rangle $, we select $\hat{\mathcal{O}}_n=\hat{P}_{i,j}\equiv 4\hat{s}_i^z\hat{s}_j^z$ \footnote{In rare situation of a decimation with  $\frac{1}{\sqrt{2}}(\vert K_+\rangle \pm \vert K_-\rangle )$, we consider as operators a combination of two $\hat{P}_{i,j}$.}. For single-site decimations, we consider the operator $\hat{\mathcal{O}}_n= 2\hat{s}_{i}^z $. The obtained average values $\mathcal{O}_n$ decay to zero in the ergodic regime, while remain close to their starting value as presented in Fig.~\ref{fig:decimate_eg}(c)(iii). In Fig.~\ref{fig:sing_dis}(a), we show a particular position configuration and decimation scheme along a selected RSRG-X branch. The initial state has a simple tensor product structure $\ket{\psi_{\mathrm{in}}}=\vert Z_{+}\rangle^{1}\otimes\vert -\rangle^{2}\otimes\vert-\rangle^{3}\otimes\vert+\rangle^{4}\otimes\vert+\rangle^{5}\otimes\vert-\rangle^{6}\otimes\vert-\rangle^{7}\otimes\vert Z_{-}\rangle^{8}$, where the superscripts represent the decimation step $n$. 
In Fig.~\ref{fig:sing_dis}(b), we present the time evolution of the operators $\hat{\mathcal{O}}_n$ and find that the expectation values are preserved except for $\hat{\mathcal{O}}_3$ and $\hat{\mathcal{O}}_5$ which reveal oscillations. The oscillations are due to a local narrow distribution of $J$s, represented by a box in Fig.~\ref{fig:sing_dis}(a). This is an example of an ergodic bubble \cite{Herviou19, Szoldra21, Szoldra24}
where local disorder is weaker than its average over the whole system. Fig.~\ref{fig:sing_dis}(c) details the time oscillations at the ergodic bubble, whose amplitude and frequency decrease with increase of $W$.

\begin{figure}
\begin{center}
\includegraphics[width=\linewidth]{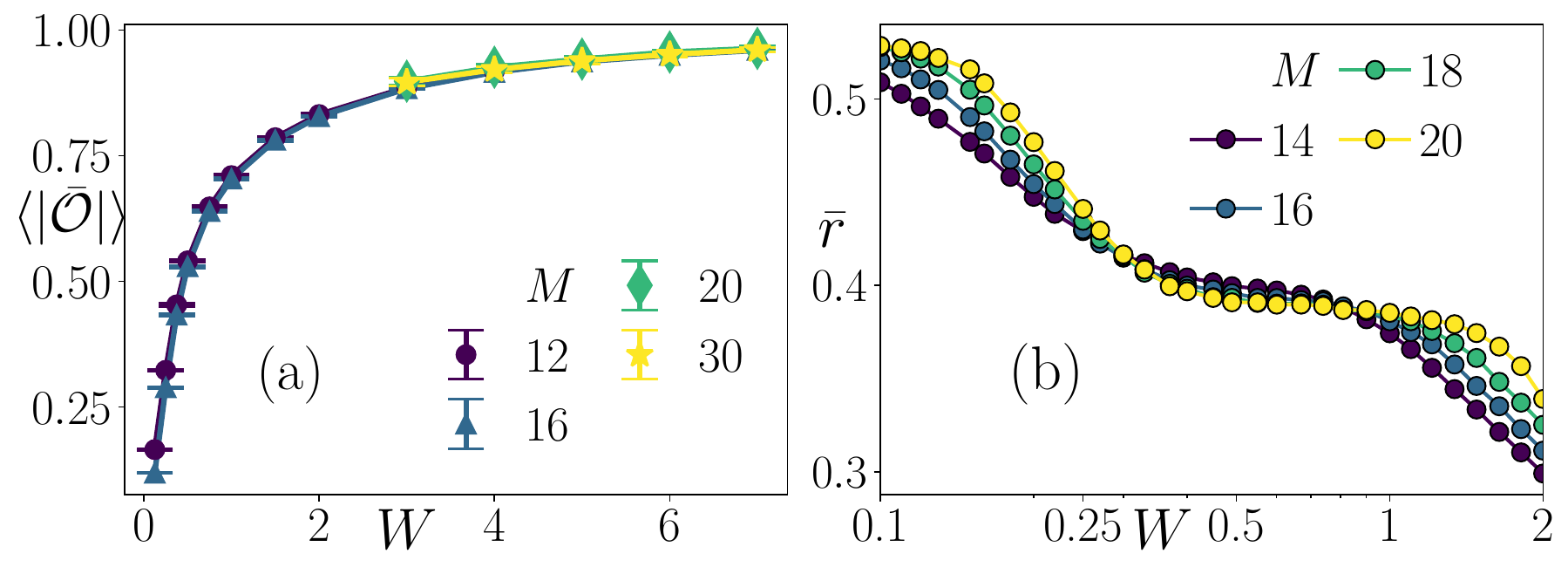}\\
\caption{(a) Long-time (up to $t=5000$) averages $\langle \vert \bar{\mathcal{O}}\vert \rangle$ of operators found in the RSRG-X procedure (see text), averaged over $10$ random RSRG-X branches and over $50$ disorder realizations. 
(b) Mean  gap ratio $\bar{r}$ as a function of disorder strength $W$ for number of spins $M$ with $\mathcal{C}=+1$}. \label{fig:op_avg}
\end{center}
\end{figure}

To further demonstrate the validity of the procedure to probe the breakdown of ergodicity in time dynamics, we consider a long-time averaged expectation value $\vert \bar{\mathcal{O}}_n \vert  = \vert \int_0^T\mathcal{O}_n(t)dt \vert $. Disorder averaged $ \langle \vert \bar{\mathcal{O}}_n \vert \rangle $ serves as ergodicity breaking indicator. As shown in Fig.~\ref{fig:op_avg}(a), $ \langle \vert \bar{\mathcal{O}}_n \vert \rangle $ decays to zero in the ergodic regime, where the mean gap ratio $\bar{r}$ (for detailed finit-size scaling analysis, see \cite{suppl_SD}) admits the GOE value $\bar{r}_{GOE}\approx 0.53$ characteristic for ergodic systems. In contrast, in the MBL regime, where $\bar{r}$ approaches (or is smaller than) the Poissonian value $\bar{r}_{PS}$, the $ \langle \vert \bar{\mathcal{O}}_n \vert \rangle $ is close to unity, showing clearly the lack of thermalization in the system.

\paragraph*{Generalization.}
Let us now consider a generic Hamiltonian of the form $H = \sum_{[\mu]} h_{[\mu]}\otimes_i \hat{s}^{\mu_i}$, where each $h_{[\mu]}$ is drawn from a certain, sufficiently broad, distribution~\cite{You16}, and $\otimes_i \hat{s}^{\mu_i}$ is a local operator. In general, the disorder \textit{is not} coupled to the on-site operators, hence the standard LIOMs are insufficient to grasp the phenomenological properties of the system at large $W$. Instead, in the absence of any special symmetry that would lead to resonances, see e.g.~\cite{Protopopov20}, the RSRG-X scheme would suffice to identify the properties of the MBL regime in the system.  Decimating the largest energy gap, a second-order perturbation theory would lead to virtual connections between the neigboring subsystems, giving rise to non-local operators conserved at large $W$. The first-order perturbation theory would fix the states of the nearby subsystems and could suppress the long-range correlations. 
\paragraph*{Conclusion.}
We studied the MBL regime arising at large disorder strengths in the bond-disordered spin-1/2 XXZ chain. The phenomenology of the MBL regime in the considered system differs significantly from models with disorder coupled to on-site operators, widely studied in the context of MBL~\cite{Abanin19, Sierant24rev}. Consequently, the standard picture of LIOMs does not describe the MBL regime in the bond-disordered XXZ spin chain, which is characterized by multimodal distribution of entanglement entropy in eigenstates, sub-Poissonian level statistics, non-trivial interplay of operators and the initial state, which has to be accounted for when probing the breakdown of ergodicity with time dynamics. The RSRG-X scheme provides a qualitative understanding of the MBL regime by approximating the highly excited eigenstates as products of domains of spins polarized along the $z$ direction and regions in which spins are paired into triplet/singlet states. The interplay between the two types of domains leads to non-trivial memory effects in the RSRG-X scheme, disallowing the identification of a global set of LIOMs for the model. We argued that the predictions of RSRG-X become quantitatively valid at strong disorder for system sizes accessible to exact diagonalization, provided that the initial distribution of couplings in the model is sufficiently broad. 

The bond-disordered XXZ spin-1/2 chain is relevant for Rydberg atoms platforms, which provide a setting to experimentally test our results. The idea of applying the RSRG-X scheme to extract phenomenological properties of MBL regime can be straightforwardly extended to other systems in which the disorder is not coupled to on-site operators. Another question involves examination of the other phenomenological features of the MBL regime in bond-disordered spin chains. For instance, the participation entropies that characterize the spread of eigenstates in the eigenbasis $\hat{s}_i^z$, see e.g.~\cite{Mace19b}, would not be vanishing even in the $W \to \infty$ limit due to the presence of triplet/singlet domains in the eigenstates. These questions are left for further studies.

\acknowledgments

The work of A.S.A. has been realized within the Opus grant
 2019/35/B/ST2/00034, financed by National Science Centre (Poland). 
P.S. acknowledges support from: ERC AdG NOQIA; MICIN/AEI (PGC2018-0910.13039/501100011033, CEX2019-000910-S/10.13039/501100011033, Plan National FIDEUA PID2019-106901GB-I00, FPI; MICIIN with funding from European Union NextGenerationEU (PRTR-C17.I1): QUANTERA MAQS PCI2019-111828-2); MCIN/AEI/ 10.13039/501100011033 and by the “European Union NextGeneration EU/PRTR"  QUANTERA DYNAMITE PCI2022-132919 within the QuantERA II Programme that has received funding from the European Union’s Horizon 2020 research and innovation programme under Grant Agreement No 101017733Proyectos de I+D+I “Retos Colaboración” QUSPIN RTC2019-007196-7); Fundació Cellex; Fundació Mir-Puig; Generalitat de Catalunya (European Social Fund FEDER and CERCA program, AGAUR Grant No. 2021 SGR 01452, QuantumCAT \ U16-011424, co-funded by ERDF Operational Program of Catalonia 2014-2020); Barcelona Supercomputing Center MareNostrum (FI-2024-1-0043); EU (PASQuanS2.1, 101113690); EU Horizon 2020 FET-OPEN OPTOlogic (Grant No 899794); EU Horizon Europe Program (Grant Agreement 101080086 — NeQST),  ICFO Internal “QuantumGaudi” project; European Union’s Horizon 2020 research and innovation program under the Marie-Skłodowska-Curie grant agreement No 101029393 (STREDCH) and No 847648  (“La Caixa” Junior Leaders fellowships ID100010434: LCF/BQ/PI19/11690013, LCF/BQ/PI20/11760031,  LCF/BQ/PR20/11770012, LCF/BQ/PR21/11840013). E.P. is supported by ``Ayuda (PRE2021-098926) financiada por MCIN/AEI/ 10.13039/501100011033 y por el FSE+".
The work of J.Z. was funded by the National Science Centre, Poland, project 2021/03/Y/ST2/00186 within the QuantERA II Programme that has received funding from the European Union Horizon 2020 research and innovation programme under Grant agreement No 101017733. A partial support by the Strategic Programme Excellence Initiative at Jagiellonian University is acknowledged. For the purpose of
Open Access, the authors applied a CC-BY public copyright
licence to any Author Accepted Manuscript (AAM) version arising from this submission.

Views and opinions expressed in this work are, however, those of the authors only and do not necessarily reflect those of the European Union, European Climate, Infrastructure and Environment Executive Agency (CINEA), nor any other granting authority. Neither the European Union nor any granting authority can be held responsible for them. No part of this work was written with the help of AI.

%


\setcounter{section}{0}
\renewcommand{\thesection}{S-\Roman{section}}
\newcommand{\snum}{S}
\renewcommand{\theequation}{\snum.\arabic{equation}}

\newpage

\section{Supplemental Material: Phenomenology of many-body localization in bond-disordered spin chains
}

\subsection{Brief Overview}
We divide the supplementary material into three broad sections. In the section \ref{sec:lrrb}, unlike in the letter, we consider a long-range random-bond XXZ model to make finite-system studies. This section includes details for generating position configurations, definitions for observables, finite-size scaling of the observables, and comparing properties with the thermodynamic limit. The following section \ref{sec:ext_RG} gives additional steps for the real space renormalization group for excited states (RSRG-X) with the calculation of trivial energy corrections, comparison of the mean gap ratio with the varying power law exponents, the illustration of thermalization, and the quantification of the spread of operators associated with the RSRG-X branch. The section \ref{sec:lev} gives insight into the sub-Poissonian behavior and quantifies the level repulsion exponent, and the section \ref{sec:sff} gives characteristics of spectral form factor of the bond disordered model. In the final section \ref{sec:num}, we provide details of the numerical methods used in the paper.

\subsection{Properties of long-range random-bond XXZ Hamiltonian\label{sec:lrrb}}

\subsubsection{Long-range random-bond XXZ model }
We consider here the random-bond XXZ Hamiltonian with long-range couplings with periodic boundary conditions, similar to \cite{Braemer22} of the form:
\begin{equation}
    \hat{\mathcal{H}}_{lr}= \frac{1}{2}\sum_{i\neq j} J_{ij}(\hat{S}_x^i\hat{S}_x^j + \hat{S}_y^i\hat{S}_y^j +\Delta \hat{S}_z^i\hat{S}_z^j).
    \label{eq::pd_Ham_lr}
\end{equation}
With the same parameters as in the letter, i.e., with $\Delta= -0.73$ and a power-law decay for both the interaction and hopping strength given as $J_{ij}=\frac{1}{\vert x_i -x_j \vert^{\alpha}}$, with $\alpha=6$. Again with a Rydberg blockade diameter $D_{rb}$, for a system with $M$ number of spins and within the length $L$, defining the spin density $\rho=D_{rb}M/L$.  We will exclusively focus on the zero-magnetization sector, and in this section, we will only consider the spin inversion symmetry sector $\mathcal{C}=+1$.
\subsection{Generation of position configuration}
\label{subsec:gen_pos}
The disorder can enter long-range random-bond XXZ Hamiltonian due to the different position configurations that can emerge with decreasing density ($\rho<1$). Here, we describe a method to generate position configurations, $\lbrace x_i \rbrace$, while avoiding the problem of R{\'e}nyi parking constant \cite{Renyi58}. Initially, shrink the Rydberg-blockaded hard spheres into points, giving a new length for the system $\tilde{L}=L-D_{rb}M$. A set of independently drawn positions is taken from the length $\tilde{L}$, $\lbrace \tilde{x}_i \rbrace $ to which $D_{rb}$ padding is added between each $\tilde{x}_i$, thus giving the original length $L$. Henceforth, we shall assume that the diameter of the Rydberg blockade $D_{rb}$ is unity; thus, the system size $L$ will exclusively control the density.

\begin{figure}[!h]
\begin{center}
\includegraphics[width=\linewidth]{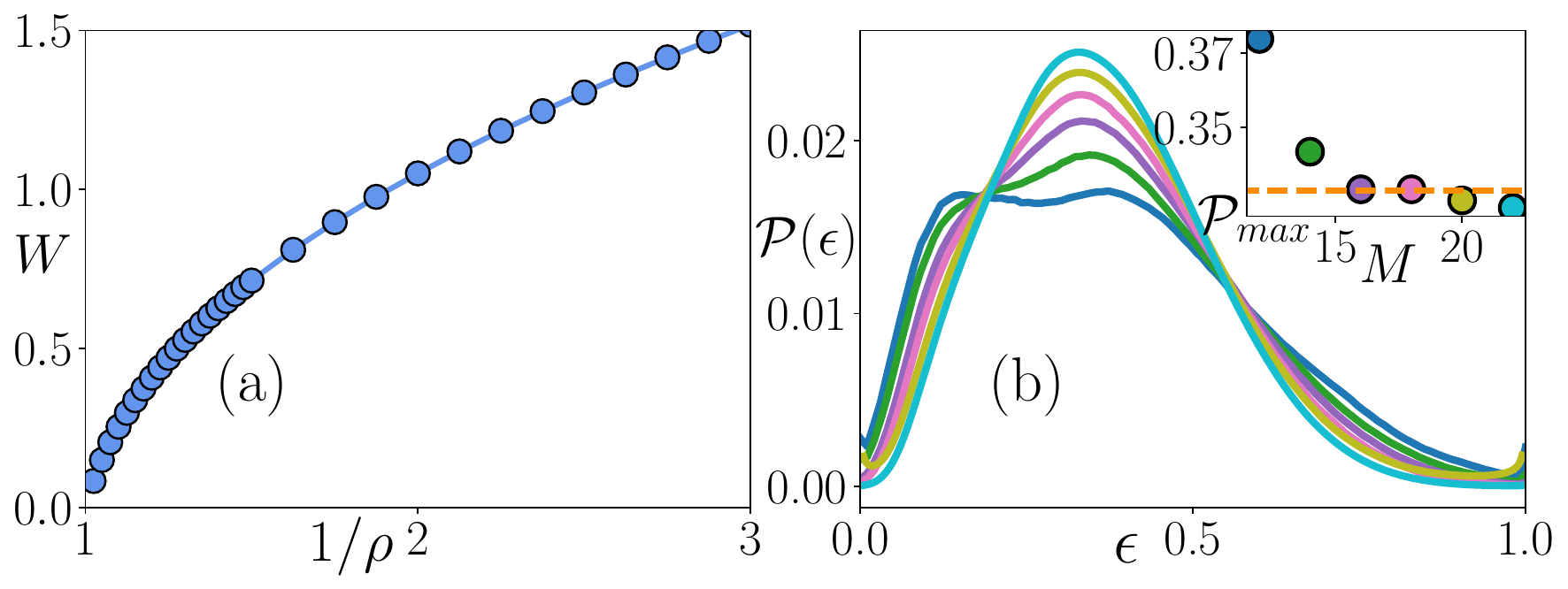}\\
\caption{(a) The disorder strength defined as in \eqref{eq::dis_str} $W$ varying with inverse density $1/\rho$. With a number of spins $M=16$ and $25000$ different position configurations. (b) The density of states (DOS) $\mathcal{P}(\epsilon)$ for the rescaled energy $\epsilon$ for different system sizes $M\in [12,22]$ for a disorder strength in the transition ($W=0.615$). The inset shows the location of the $\epsilon$ at the maximum density of states, varying with the number of spins. The dashed orange line corresponds to the $\epsilon$ value $0.333$.} \label{fig:dis_var_vs_dos}
\end{center}
\end{figure}

Fluctuations of the parameters in space characterize disorders in spin systems. In $\hat{\mathcal{H}}_{lr}$, the disorder occurs as fluctuations in the parameter $J_{i,j}$, with both the fluctuation and the mean value around which the fluctuation occurs dependent on the density. Thus, defining disorder strength is natural as a fluctuation around the effective $J$ term as  
\begin{equation}
    W=\frac{\sigma_{ J^i_{eff} }}{\langle J^i_{eff} \rangle} 
    \label{eq::dis_str}
\end{equation}
wherein, $J_{eff}^i$ is $\sum_jJ_{i,j}$ and $\sigma$ represents standard deviation. The disorder strength increases monotonically with decreasing density, as shown in Fig. \ref{fig:dis_var_vs_dos}(a). Henceforth, we shall refer to a particular position configuration as a disorder realization.

\subsubsection{Observables}
To depict the crossover between the ergodic and MBL regimes, we consider the mean gap ratio $\bar{r}$\cite{Oganesyan07} as in the letter and the average rescaled half-chain entanglement entropy (EE) $S$. The half-chain EE is defined as the von Neumann entropy of the reduced density matrix $\rho_{M/2}$:
\begin{equation}
    S_{M/2}=-Tr[\rho_{M/2}\ln \rho_{M/2}]
\end{equation}
where $\rho_{M/2} = Tr_{1,2,\cdots,M/2}\vert \Psi\rangle \langle \Psi \vert $ is obtained by tracing out half of the system. To minimize the dependence of the system size for EE, we rescale it by the corresponding random matrix theory value ${S}_{RMT}=(M/2)\ln(2)+[1/2+\ln(2)]/2-1/2$ as $S_{\Psi}=S_{M/2}/S_{RMT}$ \cite{Vidmar17}.
Finally, similar to that of $\bar{r}$, we average the rescaled half-chain  EE over the eigenstates corresponding to a particular disorder realization within an eigenvalue range and then over different disorder realizations, $S$.

\begin{figure}%
\includegraphics[width=\linewidth]{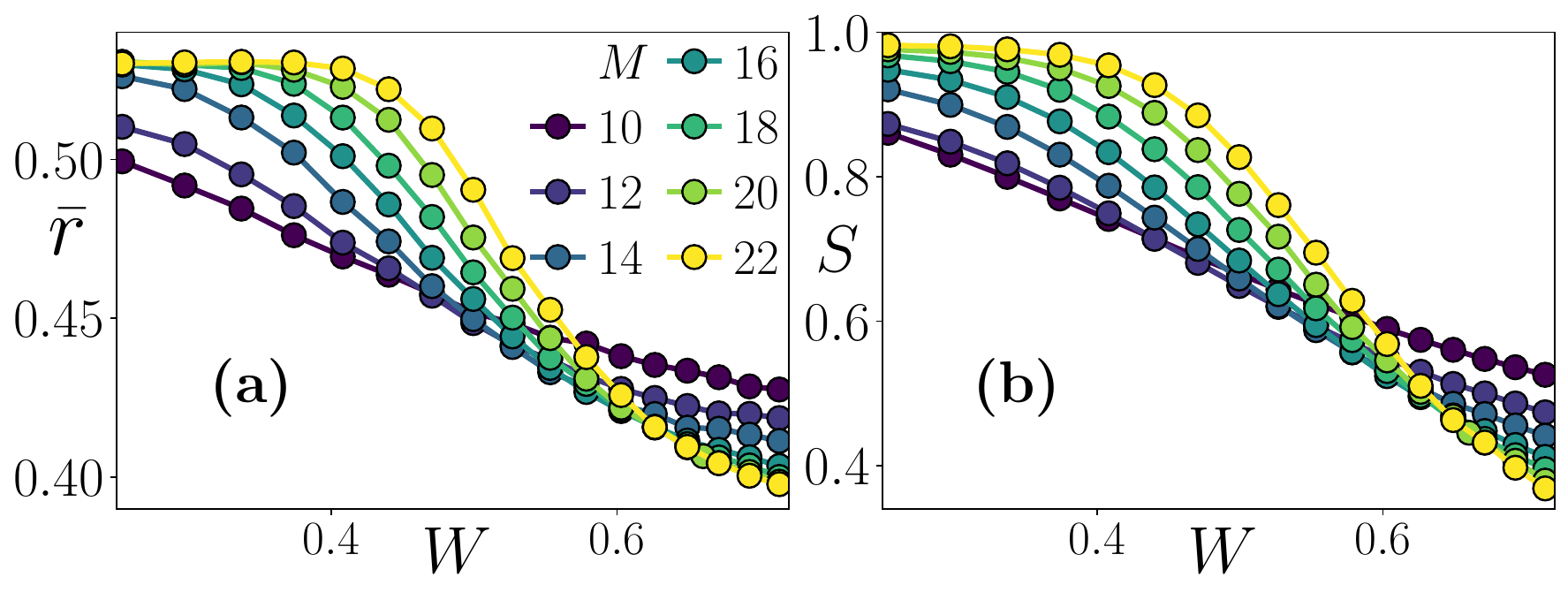}
\caption{The dependence of (a) the mean gap ratio $\bar{r}$, (b) the rescaled average half-chain EE ${S}$, with varying strengths $W$ for different number of spins $M \in [10, 22]$.}
\label{fig:basicplots}
\end{figure}

For each disorder strength $W$ and the number of spins $M$, the observables are calculated by averaging over eigenstates near the rescaled energy
\begin{equation}
    \epsilon = \frac{E-E_{min}}{E_{max}-E_{min}} = 0.333
\end{equation}
for a given disorder realization and subsequently over different disorder realizations. Note that we do not consider the typical rescaled energy of $\epsilon = 0.5$ to average the observables here, since the density of states (DOS) peaks around $\epsilon \approx 0.333$ for higher number of spins ($M\geq 14$) for the disorder strength of interest to us, as evident in Fig. \ref{fig:dis_var_vs_dos}(b), this is a result of the smaller energy gap between $\vert Z_{\pm} \rangle $ and $\vert - \rangle $. The number of eigenvalues and positional configurations used in this study for different $M$s and the corresponding Hilbert space dimension $\mathcal{N}$ considered are enumerated in Table \ref{table:disord}. The mean gap ratio and the average rescaled half-chain EE dependence on disorder strength and number of spins are shown in Fig. \ref{fig:basicplots}. 

\begin{table}[htb]
\centering
\caption{Table showing the approximate number of states considered and the minimum number of disorder realizations used for the statistical observables.}
\begin{ruledtabular}
\begin{tabular}{c c c c} 
 $M$ & $\mathcal{N}$ & No. eigenstates & No. realizations \\ [0.5ex] 
 \colrule
 $10$ & $126$ & $3$ & $2\cdot 10^5$  \\
 $12$ & $462$ & $5$ & $10^6$\\
 $14$ & $1716$ & $17$ & $10^5$\\
 $16$ & $6435$ & $64$ & $8\cdot 10^3$ \\
 $18$ & $24310$ & $243$ & $2\cdot 10^3$ \\
 $20$ & $92378$ & $10^3$ & $10^3$\\
 $22$ & $352716$ & $10^3$ & $700$\\[1ex] 
\end{tabular}
\end{ruledtabular}
\label{table:disord}
\end{table}

\subsubsection{Finite-size scaling analysis}
In the case of a phase transition, the correlation length $\xi$ diverges near the critical disorder strength and acts as the dominant length scale. As a result, the normalized observable, $X$, takes a functional form 
\begin{equation}
    X=\mathcal{G}(L/\xi),
\end{equation}
where $\mathcal{G}(.)$ is a continuous function. 

The two prominent proposals for how the correlation length $\xi$ diverges in the MBL transition is either as a standard second-order phase transition where $\xi$ diverges as a power-law fashion $\xi_0=\frac{1}{\vert W-W^*\vert^\nu }$, $\nu$ is the critical exponent and $W^*$ is the critical disorder strength based on the early renormalization group approaches \cite{Vosk15,Potter15}, or assuming a Berezinskii-Kosterlitz-Thouless (BKT) like divergence: $\xi_{B}=\exp\left\lbrace \frac{b_\pm}{\sqrt{\vert W-W^* \vert }}\right\rbrace$, where $b_{\pm}$ are parameters on the two sides of the transition, based on the more recent real-space renormalization group approaches \cite{Goremykina19,Dumitrescu19,Morningstar19,Morningstar20} with the leading mechanism as the avalanche scenario of delocalization in the MBL phase. 

To best approximate the correlation length parameters, we follow the cost function minimization method introduced by Suntajs et al. \cite{Suntajs20,Suntajs20e} for the XXZ model in the presence of random disorder. This method has been used for other MBL models \cite{Aramthottil21,Yousefjani23}; recent studies have also shown that it predicts the critical point for the 3-D Anderson \cite{Suntajs21} and the zero-dimensional quantum sun model \cite{Suntajs22,Pawlik23} accurately. 
\begin{table}[htb]
\centering
\caption{ Cost function $\mathcal{C}_X$ comparison for finite-size scaling
with number of spins dependent drifts in the critical disorder strength.}
\begin{ruledtabular}
\begin{tabular}{c c c c} 
& $\mathcal{C}_{\bar{r}}[\xi_0]$ & $\mathcal{C}_{\bar{r}}[\xi_{B}(b_+=b_-)]$ & $\mathcal{C}_{\bar{r}}[\xi_{B}(b_+\neq b_-)]$ \\ [0.5ex] 
 \colrule
 $W_0$ & $1.1358$& $1.566$& $1.226$\\
 $W_0+W_1M$ & $0.312$& $0.226$& $0.222$\\
 $W_0+W_1\ln (M)$ & $0.304$& $\textbf{0.183}$& $\textbf{0.183}$\\
 $W_0+W_1/{M}$& $0.686$& $0.462$& $0.462$\\
 $W_0+{W_1}/{[\ln(M)]}$ & $0.336$& ${0.199}$& $0.199$\\
 $W_0+W_1M^{\gamma}$ & $0.304$& $\textbf{0.140}$& $\textbf{0.140}$\\
  $W(M)$ & $0.282$& $\textbf{0.106}$& $\textbf{0.106}$\\[0.5ex] 
\end{tabular}
\end{ruledtabular}
\begin{ruledtabular}
\begin{tabular}{c c c c} 
& $\mathcal{C}_S[\xi_0]$ & $\mathcal{C}_S[\xi_{B}(b_+=b_-)]$ & $\mathcal{C}_S[\xi_{B}(b_+\neq b_-)]$ \\ [0.5ex] 
 \colrule
 $W_0$  & $1.117$& $1.338$& $1.211$\\
 $W_0+W_1M$ & $0.122$& $0.0.279$& $0.168$\\
 $W_0+W_1\ln (M)$ & $0.102$& $0.131$& $0.098$\\
 $W_0+W_1/{M}$ & $0.397$& $0.542$& $0.499$\\
 $W_0+{W_1}/{[\ln(M)]}$& $0.109$& $\textbf{0.0504}$& $\textbf{0.0504}$\\
 $W_0+W_1M^{\gamma}$ & $0.102$& $\textbf{0.0507}$& $\textbf{0.0507}$\\
  $W(M)$ & $0.086$& $\textbf{0.030}$& $\textbf{0.030}$\\[0.5ex] 
\end{tabular}
\end{ruledtabular}
\label{table:CF}
\end{table}

Suppose that the observables of interest are continuous monotonic functions of $sgn[W-W^*]L/\xi$. The cost function for a quantity $X=\lbrace X_j\rbrace$ that consists of $N$ different $W$ and $L$ is defined as 
\begin{equation}
    \mathcal{C}_X=\frac{\sum_{j=1}^{N-1}\vert X_{j+1}-X_j\vert }{\max\lbrace X_j \rbrace-\min\lbrace X_j \rbrace}
\end{equation}
where $X_j$'s are sorted according to non-decreasing values of $sgn[W-W^*]L/\xi$. For an ideal collapse with $X$ being a monotonic function of $sgn[W-W^*]L/\xi$, we must have  
$\sum_j \vert X_{j+1}-X_j\vert=\max\lbrace X_j\rbrace-\min\lbrace X_j\rbrace$,
and thus  $\mathcal{C}_X=0$. Thus, the best collapse corresponds to the global minima of $\mathcal{C}_X$ for different correlation lengths and functional forms of $W^*(L)$. The final section details the numerical parameters for the cost function minimization performed.

To perform an unbiased finite size scaling analysis for $\hat{\mathcal{H}}_{lr}$, we consider observables of system sizes $M\in [12,22]$ within disorder strengths $W\in [0.3,0.725]$, minimizing the cost function for the correlation lengths $\xi_0,\xi_B$ with different functional forms of the critical disorder, $W^*$ on the number of spins ($M$) as shown in Table \ref{table:CF}.

With a fixed critical disorder strength $W^{*}=W_0$, the power law correlation length furnishes a smaller value $\mathcal{C}_X$ for both the mean gap ratio, $\bar{r}$, and the average rescaled half-chain EE, $S$. Finding critical disorder strengths as $W^*_r=0.584$ and $W^*_S=0.607$ and critical exponents $\nu_{\bar{r}}=0.661$ and $\nu_{S}=0.662$.

\begin{figure}%
\includegraphics[width=\linewidth]{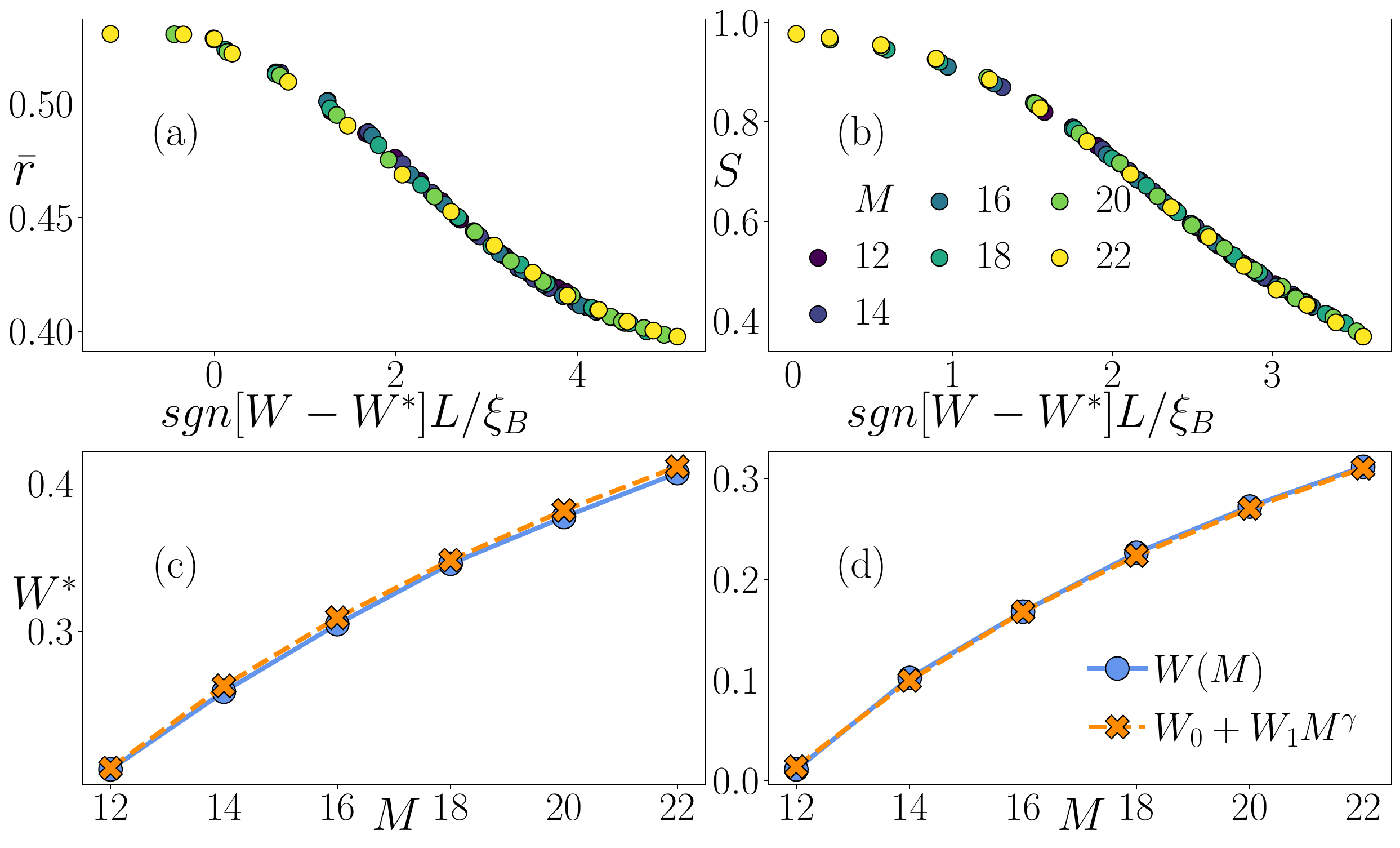}
\caption{Finite-size scaling for mean gap-ratio ($\bar{r}$) and rescaled averaged half-chain EE ($S$), with correlation length $\xi_B$ and the critical disorder strength with a drift of functional form $W^*=W_0+W_1M^{\gamma}$ is shown in  (a) and (b). While (c) and (d) show the dependency of free disorder strength $W^*(M)$, on $M$ in comparison with the functional form $W^*=W_0+W_1M^{\gamma}$. }
\label{fig:Cf}
\end{figure}

While assuming different functional forms for the critical disorder strength, the best-minimized cost function value is a BKT-like correlation with sublinear scaling, Table \ref{table:CF}; a functional form $W^*=W_0+W_1L^{\gamma}$ has the best cost function value for both $\bar{r}$ and $S$. The minimized values for the drift $(W_0,\gamma)$ are quite different for $\bar{r}\rightarrow(1.36,-0.32)$ and $S\rightarrow(1.00,-0.59)$. With the nonuniversal parameter, $b$, for $\bar{r}$ as $0.804$ and for $S$ as $1.09$. There appears to be no improvement in $\mathcal{C}_X$ when considering different nonuniversal parameters $b_+ \neq b_-$, as the observables considered appear mostly in the localized regime. Considering a generic function for $W^*=W(M)$, where each $W(M)$ is chosen individually for $M$, we find the functional form $W^*=W_0+W_1L^{\gamma}$ to approximate $W(M)$ as shown in Fig. \ref{fig:Cf}.

These results, though, do not indicate a strong interpretation of $W_0$ in the functional form $W^*=W_0+W_1L^{\gamma}$, as the critical disorder strength at the thermodynamic limit as in the 3-D Anderson model \cite{Suntajs21}, this is because the values found $W^*$ for the considered number of spins $M$ are pretty far from $W_0$. However, $2\gamma_{\bar{r}} \sim \gamma_{S}$ indicates that $S$ might more evidently show properties of the thermodynamic limit in the $M$ considered than in $\bar{r}$.

\subsubsection{How well estimated are the finite-spin observables compared to the thermodynamic limit?}

To address the question of how well estimated the finite-spin observables are to properties in the thermodynamic limit, we define two disorder strengths as in \cite{Sierant21constrained,Sierant23}
\begin{enumerate}
    \item $W_T^X(M)$: the disorder strength at which, for a given number of spins $M$, the quantity $X$ deviates from a percentage of its ergodic value compared to the localized non-ergodic value.
    \item $W_{\times}^X(M)$: the disorder strength at which the curves $X(W)$ cross for systems with the number of spins $M-1$ and $M+1$.
\end{enumerate}

The disorder strength $W_T^X(M)$ can be considered as the boundary of the ergodic regime. In contrast, the disorder strength $W_{\times}^X(M)$ when $M \rightarrow \infty$ approximates the critical disorder strength assuming MBL persists in the thermodynamic limit, $W_{\infty}$. Depending on the type of transition, the regime between $W_T^X(M)$ and $W_{\times}^X(M)$ either vanishes for $L\rightarrow\infty$ or tends to a constant.

For the random disordered XXZ model, the disorder strength $W_T^X(M)$ increases monotonously with the number of spins, $W_T^X(M)\sim  M$ and $W_{\times}^X(M)\sim W_{\infty}-a_1/L$ \cite{Sierant20p}. However, these scalings are incompatible with the existence of MBL in the large length limit when $W_T^X(M)$ exceeds $W_{\times}^X(M)$ at $M_0^{XXZ}\approx 50$.
Thus, approaching the number of spins $M_0^{XXZ}$ indicates the presence or absence of MBL and can be extended to compare how well-approximated observables are with the thermodynamic limit.

\begin{figure}[!h]
\begin{center}
\includegraphics[width=\linewidth]{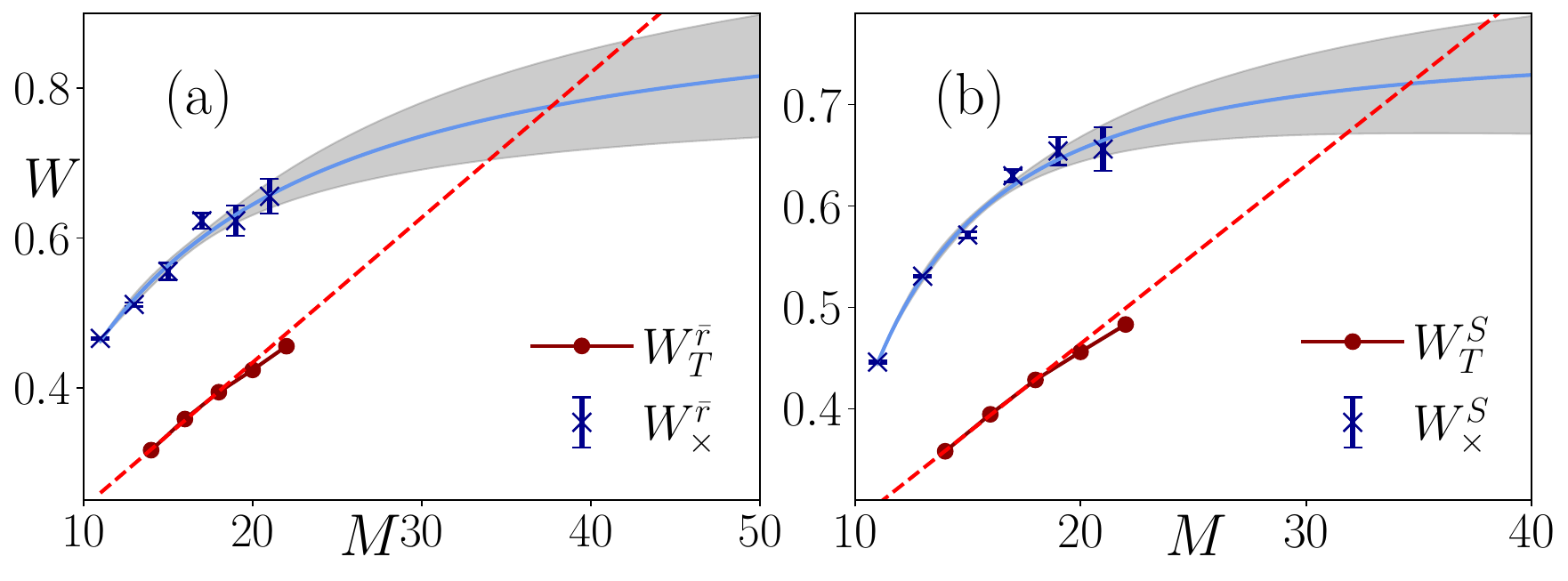}\\
\caption{(a) The crossing point disorder strength $W_{\times}^{\bar{r}}$ and the near ergodic disorder strength $W_T^{\bar{r}}$ varying with $M$ for $\bar{r}$. The bold blue line denotes a fit on $W_{\times}^{\bar{r}}$ with a functional form $W_{\infty}+a_1/M+a_2/M^2$ the grayed area denotes the associated errors, while the dashed red line denotes a linear fit for $W_T^{\bar{r}}$. In (b), the same results as in (a) are shown for $S$.}\label{fig:W_cr}
\end{center}
\end{figure}

In case of $\hat{\mathcal{H}}_{lr}$, the results of which are shown in Fig. \ref{fig:W_cr}, with $W_{\times}^X(M)$ we consider a functional form $W_{\times}^X(M)=W_{\infty}+a_1/M+a_2/M^2$ from which $W_{\infty}$ is extracted for $\bar{r}$ and $S$, respectively, as $W_{\infty}^{\bar{r}}=0.951\pm 0.157$ and $W_{\infty}^{S}=0.764\pm 0.134$. A substantial percentage difference is considered from the ergodic value to estimate $W_{T}^X(M)$ because, for $M \leq 14$, $X$ does not saturate to the ergodic value even with small disorder strengths,  $8.5\%$ for $\bar{r}$ and $20\%$ for $S$. Similar to the Floquet model considered in \cite{Sierant23} $M \geq 20$, $W_{T}^X(M)$ deviates from a linear function, so we make a linear fit with $M \geq 20$ to make a comparison with the XXZ model. From the curves extracted from $W_{T}^X(M)$ and $W_{\times}^X(M)$ we find that the value $M$ at which the two curves cross as $M_0^{\bar{r}}\in[34.5,43.1]$ and $M_0^{S}\in[31.9,38.1]$ is less than the XXZ model with random disorder \cite{Sierant20p} hinting at better approximating thermodynamic properties, but higher than the Floquet model in \cite{Sierant23}. 

\subsection{Extension of RSRG-X results\label{sec:ext_RG}}
\subsubsection{Random-bond XXZ model}
For this section, \ref{sec:lev} and \ref{sec:sff} we define the random-bond XXZ model with nearest-neighbor interaction as in the letter again for ease of reading.
\begin{equation}
    \hat{\mathcal{H}}= \sum_{i=1}^{M-1} J_{i,i+1}(\hat{s}_i^x\hat{s}_{i+1}^x + \hat{s}_i^y\hat{s}_{i+1}^y +\Delta \hat{s}_i^z\hat{s}_{i+1}^z). 
    \label{eq::Ham}
\end{equation}
The parameters are the same as in \eqref{eq::Ham}. The disorder strength is defined $W\equiv 1/\rho -1$ as in the letter.
\subsection{Second order perturbation for $\vert Z_{\pm} \rangle$ states. \label{Z}}
Fixing spins to a state $\vert Z_{\pm} \rangle$ at a decimation leads to effective nearby operators modified by a first-order perturbation to the chemical potential form ${\hat{\mathcal{H}}'}_{i,j}$ thereby typically leading to a process of growing domain spins.
There is an exception when the generation of domain spins leads to degenerate states of the form $\vert K_+ \rangle \equiv \vert \uparrow \rangle\vert Z_- \rangle \vert \uparrow \rangle $, $\vert K_- \rangle \equiv \vert \downarrow \rangle\vert Z_+ \rangle \vert \downarrow \rangle $. The states $\vert K_{\pm} \rangle $ could be coupled through second-order perturbation if the resulting energy gap is greater than the effective neighbouring operators ${\hat{\mathcal{H}}'}_{i,j}$ to $\vert K_{\pm} \rangle $. Eventually, this coupling would result in operators of the form $\hat{\mathcal{H}}_{i,j}$.

\begin{figure}%
\includegraphics[width=\linewidth]{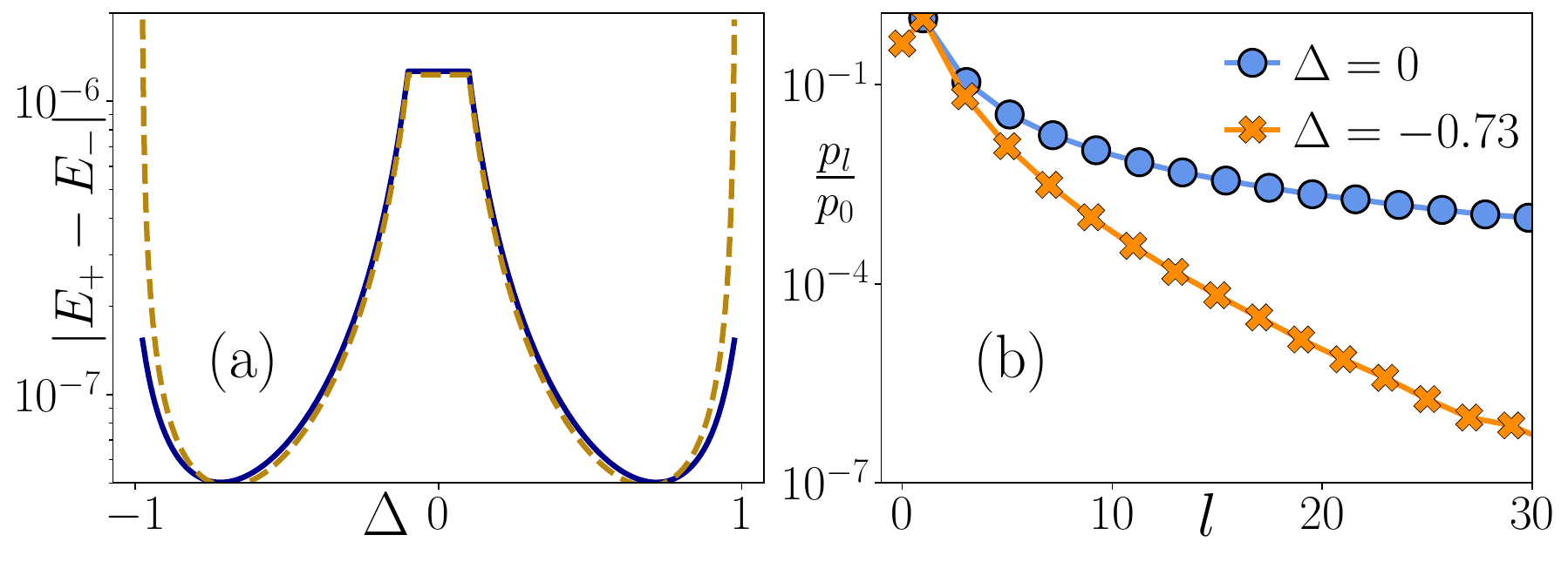}
\caption{(a) Show the absolute value of energy difference between the approximate states of $\vert \phi_+\rangle $ and $\vert \phi_-\rangle $, given as $\vert E_+ - E_- \vert$ represented with bold line, while dashed lines represents $\tilde{J}_{1,6}^+$ from \eqref{eq::J_16}. (b) The probability of spread of operators, $p_l$, rescaled with $p_1$, where $l$ represents the distance between sites. The rescaled values shown in the above figure were calculated for $10$ different decimations $\gamma$s, with $10^4 $ positional configurations for $M=100$ with $\mathcal{C}=+1$ .}
\label{fig:RSRG_ext}
\end{figure}

Consider the strongest bond is between the spins $i=3$ and $i=4$. Fixing the spins with the choice $\vert Z_{\pm}\rangle $ results in operators of the form $\hat{\mathcal{H}}'_{4,5}$ , $\hat{\mathcal{H}}'_{2,3}$ these operators in further decimation steps can give states $\vert K_{\pm} \rangle$ and generation of operators $\hat{\mathcal{H}}'_{5,6}$ and $\hat{\mathcal{H}}'_{1,2}$. In such a scenario, in addition to the operators $\hat{\mathcal{H}}'_{5,6}$, $\hat{\mathcal{H}}'_{1,2}$, a second-order perturbation couples the states $\vert K_{\pm} \rangle $. To elucidate this coupling better, assume that the spins  $i=3$ , $i=4$  are instead fixed with the states  $\frac{1}{\sqrt{2}}(\vert Z_{+} \rangle \pm \vert Z_{-} \rangle )$ which would result in an effective operator  of the form
\begin{eqnarray}
\hat{\mathcal{H}}''_{2,5} &=& \frac{J'}{2}\sum_{\Theta=\pm 1}\biggl[\frac{\pm\Theta}{(1+\Theta\Delta_{3,4})}(\hat{s}^+_2\hat{s}^+_5+h.c.)\nonumber\\
&~&-\frac{1}{(1+\Theta\Delta_{3,4})}(\hat{s}_2^+\hat{s}_{5}^- +h.c. )\biggl]
\label{eq::K}
\end{eqnarray}
where, $J'$ is defined as $J'\equiv\frac{J_{2,3}J_{4,5}}{2J_{3,4}}$. The second term in \eqref{eq::K} is just the operator considered with second-order perturbation with choice of states $\vert Z_{\pm} \rangle $. The first term, which vanishes when $\vert \Delta \vert \rightarrow 0$, results in states $\frac{1}{\sqrt{2}}(\vert Z_{+} \rangle \pm \vert Z_{-} \rangle )$ on spins $i=2$, $i=5$. Combined with restrictions due to $\hat{\mathcal{H}}'_{4,5}$ and $\hat{\mathcal{H}}'_{2,3}$ and local $U(1)$ symmetry, leads to states of the form $\frac{1}{\sqrt{2}}(\vert K_+ \rangle \pm \vert K_- \rangle )$ . If the energy gap resulting from the effective neighbour operators $\hat{\mathcal{H}}'_{5,6}$, $\hat{\mathcal{H}}'_{1,2}$, are smaller than $\hat{\mathcal{H}}''_{2,5}$ we would make a decimation fixing to states $\frac{1}{\sqrt{2}}(\vert K_+ \rangle \pm \vert K_- \rangle )$, which would result in modifying the nearby operators via second-order perturbation to take the form $\hat{\mathcal{H}}_{1,6}$ with parameters
\begin{equation}
     \tilde{J}_{1,6}^{\pm}=\sum_{\Theta=\pm 1}\frac{J_{1,2}J_{5,6}}{2J'\frac{1}{1\mp\Theta\Delta_{3,4}}+\Theta(J_{2,3}\Delta_{2,3}+J_{4,5}\Delta_{4,5})}
     \label{eq::J_16}
\end{equation}
and
\begin{equation}
\tilde{J}_{1,6}^{\pm}\tilde{\Delta}_{1,6}^{\pm}=\frac{1}{4}\left(\frac{J_{1,2}\Delta_{1,2}J_{5,6}\Delta_{5,6}}{\pm J'(\frac{1}{1+\Delta_{3,4}}-\frac{1}{1-\Delta_{3,4}})}\right).
\end{equation}
To check the validity of \eqref{eq::J_16} in Fig. \ref{fig:RSRG_ext}(a) we consider a particular disorder realization that satisfies the above condition for a system with $M=6$  and make a comparison between $\tilde{J}^+_{1,6}$ and the absolute energy difference $\vert E_+ - E_- \vert$ between two states with maximum overlap to the states $\vert \phi_{\pm} \rangle \equiv \frac{1}{\sqrt{2}}(\vert K_+ \rangle + \vert K_- \rangle )\vert \pm \rangle_{1,6}$. We find that with varying $\Delta$, the absolute energy difference is well approximated by $\tilde{J}^+_{1,6}$  except when $\vert \Delta\vert \rightarrow 1 $.

\subsubsection{Trivial energy corrections.}
 
At each decimation step, one encounters trivial energy corrections $\delta E$, which are scalar values that modify the energy. These corrections are essential for the precise calculation of energy; in Table \ref{table:trivial_ener}, we show the trivial energy correction associated with each decimation step. Where, $J'$ is  again defined as $J'\equiv\frac{J_{2,3}J_{4,5}}{2J_{3,4}}$.

\begin{table}[ht]
    \centering
    \caption{Trivial energy $\delta E$ corrections with second order perturbation for each choice of state.}
    \begin{ruledtabular}
        \begin{tabular}{cc} 
            State & $\delta E$ \\ 
            \hline
            $ \vert \pm \rangle $ & $\mp\frac{J_{2,3}^2+J_{4,5}^2}{4J_{3,4}}\biggl(\frac{\Delta_{3,4}^2}{4}+\frac{1}{1\pm\Delta_{3,4}}\biggl) $\\
            $ \frac{1}{\sqrt{2}}(\vert Z_{+} \rangle\pm\vert Z_{-} \rangle)$ & $\frac{\left(J_{2,3}^2+J_{4,5}^2\right)}{8J_{3,4}}\left(\frac{1}{(1+\Delta_{3,4})}-\frac{1}{(1-\Delta_{3,4})}\right)$\\
            $ \vert K_{\pm} \rangle $ & $\frac{J_{2,3}^2+J_{4,5}^2}{-4\left(J_{2,3}\Delta_{2,3}+J_{4,5}\Delta_{4,5}\right)+8J'\left(\frac{1}{1\pm\Delta_{2,3}}\right)}$
            \\ 
            $ $ & $-\frac{J_{2,3}^2+J_{4,5}^2}{4\left(J_{2,3}\Delta_{2,3}+J_{4,5}\Delta_{4,5}\right)+8J'\left(\frac{1}{1\mp\Delta_{2,3}}\right)}$
            \\ 
            $ $ & $\pm\frac{J_{2,3}^2\Delta_{2,3}^2+J_{4,5}^2\Delta_{4,5}^2}{8J'\left(\frac{1}{1+\Delta_{2,3}}-\frac{1}{1-\Delta_{2,3}}\right)}$
        \end{tabular}
    \end{ruledtabular}
    \label{table:trivial_ener}
\end{table}

\subsection{Mean gap ratio with broader distribution}
\begin{figure}%
\includegraphics[width=\linewidth]{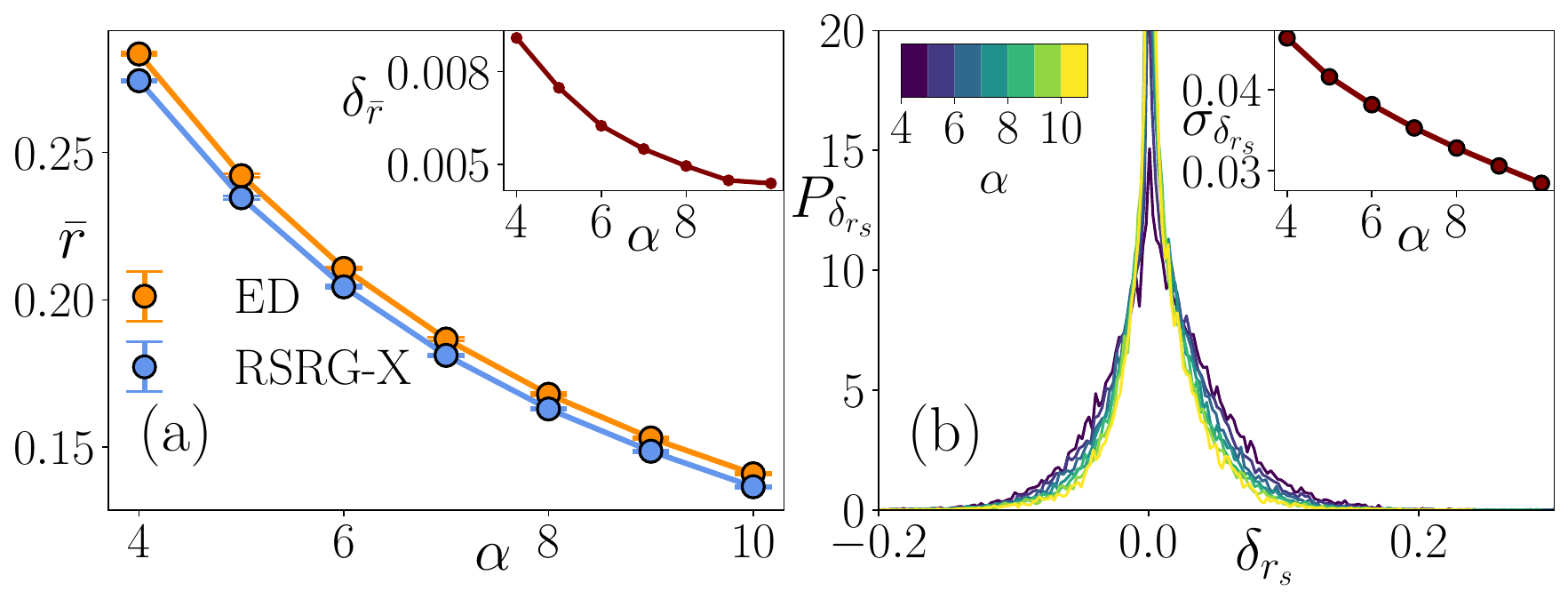}
\caption{(a) The mean gap ratio ($\bar{r}$) obtained by RSRG-X and exact diagonalization (ED) varying with the exponent $\alpha$. The inset shows the relative difference $\delta_{\bar{r}}=\bar{r}_{ED}-\bar{r}_{RG}$. (b) Distribution of relative differences in the gap ratios averaged over a disorder realization ($P_{\delta{r_s}}$) with the exponent $\alpha$. The inset shows the variation of the standard deviations of $P_{\delta{r_s}}$ with respect to $\alpha$, given as $\sigma_{\delta_{r_s}}$. The system size considered is $M=10$ in the $\mathcal{C}=+1$ sector with fixed disorder strength $W=8.0$, with $25000$ disorder realizations and considering all the eigenvalues }
\label{fig:var_alpha}
\end{figure}

On considering exponents for the power law controlling $J$'s, $\alpha$, we find that RSRG-X well approximates the mean gap ratio $\bar{r}$ compared to the exact diagonalization results, Fig. \ref{fig:var_alpha} (a). More significantly, the distribution of the gap ratio averaged over a disorder realization $P_{\delta_{r_s}}$ decreases with $\alpha$ strongly Fig. \ref{fig:var_alpha} (b).

\subsubsection{Illutration of thermalization}
\begin{figure}
\begin{center}
\includegraphics[width=\linewidth]{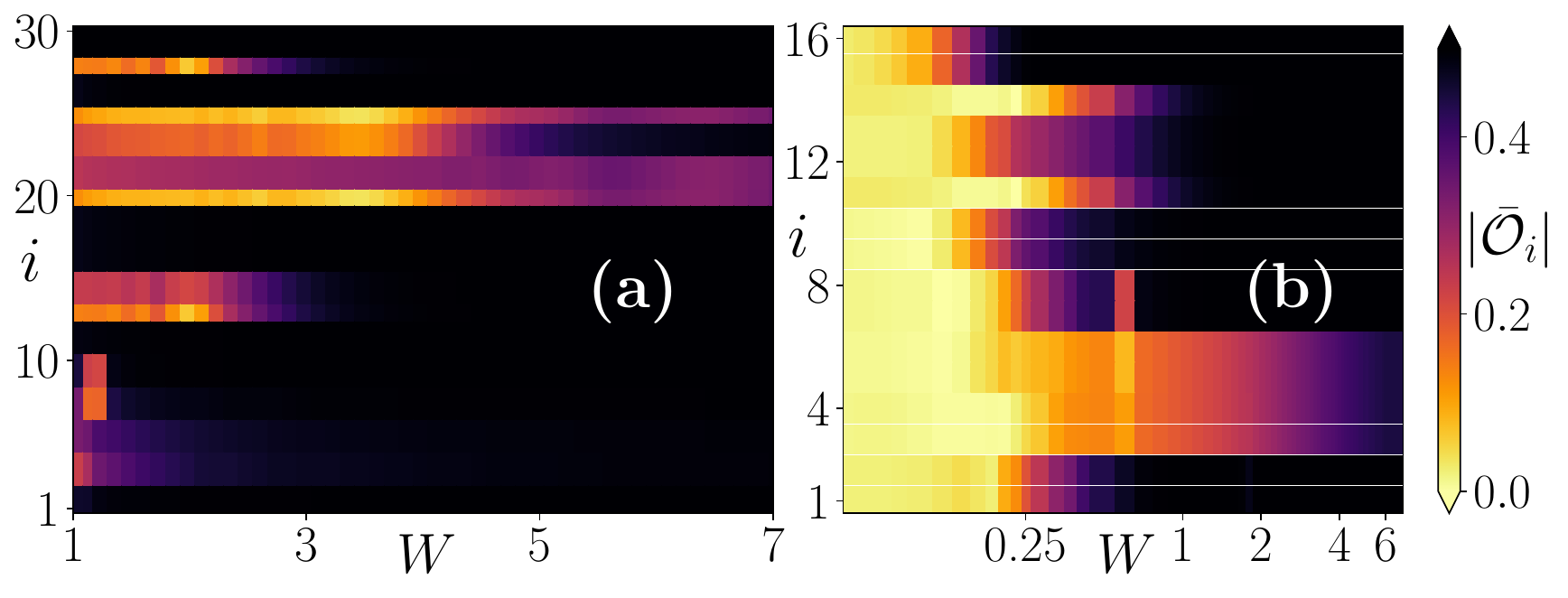}\\
\caption{The deviation of the absolute value of  long-time averaged operators $ \vert \bar{O_i}\vert $
at each site with varying densities. (a)  $M=30$,  (b) $M=16$  for the same positional configuration and set of operators as in Fig. (3) in the letter. In both cases averages are taken up to time $t=5000$.} \label{fig:therm}
\end{center}
\end{figure}
In \cite{Chandran15}, infinite time averages of local operators are taken, giving the correlation length $\zeta$ associated with the local integral of motions  (LIOMs). Here, we look for the absolute value of infinite time averages of the operators corresponding to an RSRG-X branch $ \vert \bar{\mathcal{O}}_n \vert  =  \vert \int_0^T\mathcal{O}_n(t)dt \vert  $, an unchanging value will indicate that RSRG-X branch is stable. A variation from the stability that does not spread through the system, which typically arises from ergodic bubbles, indicates that $ \hat{\mathcal{O}}_n$ are not the best form of local operators in the given RSRG-X branch. However, the system as a whole is still localized. While the operators' instability throughout the system would indicate thermalization. To visualize these operators spatially, in Fig. \ref{fig:therm} we consider that the values $\vert \bar{\mathcal{O}}_n \vert $ are distributed equally in space, $\vert \bar{\mathcal{O}}_i \vert = \frac{1}{2}\vert (\bar{\mathcal{O}}_n^{i,j}) \vert $.  Here $i,j$ denotes indices associated of spins.
 Fig.~\ref{fig:therm}(a) considers a position configuration with $M=30$ for a strong disorder with a particular disorder realization and RSRG-X branch. In contrast, in Fig.~\ref{fig:therm}(b), we consider a smaller system size with the same position configuration and operators as in Fig. (3) of the letter to explore weaker disorder regimes. In  Fig.~\ref{fig:therm}(a), we find that at strong disorder, most of the operators are preserved, while with decreasing disorder strength, new ergodic bubbles pop up as seen in the region between sites $12-15$ but stay strictly local in most cases except between sites $3-6$ at disorder strength $W=1$. While in Fig.~\ref{fig:therm}(b), we find as we approach the region of weak disorder $W\sim 0.1$, the ergodic bubble between sites $3-6$ starts to thermalize the nearby region, eventually thermalizing operators at all sites indicating ergodicity.
\subsection{Localization length}

To contrast the spread of operators associated with an RSRG-X branch $\hat{\mathcal{O}}_{n,k}^{i_a,i_b}$ with the correlation length of LIOMs with on-site disodered systems \cite{Serbyn13b,Chandran15,Ros15} we look for the spread of operators on the spins. Here, $i_a$ denotes the spin indices and $k$ denotes the associated eigen index, by defining the distance between the spins of these operators as $l=\vert i_a-i_b \vert$, for a single site operator $l=0$, we define a probability of the spread as  
\begin{equation}
    p_l=\frac{1}{\eta}\sum_{n,k}\vert\vert \hat{\mathcal{O}}_{n,k}^{l}\vert \vert 
    \label{eq::eff_LIOM}
\end{equation}
Where $ \vert\vert \hat{\mathcal{O}}_{n,k}^{l}\vert \vert  $ denotes the Frobenius norm of the operator and $\eta$ is a normalization constant such that $\sum_l p_l = 1$. If the probability distributions are exponentially distributed, $p_l\propto \exp{(-l/\zeta_0)}$, where $\zeta_0$ is the length of the localization at  asymptotic disorder strength, we would identify our system as localized.

In Fig.~\ref{fig:RSRG_ext}(b), using RSRG-X and assuming the Frobenius norm associated with the operator of an RSRG-X branch as unity, we compare $p_l$ for a finite $\Delta$ with that of $\Delta = 0$. For $\Delta=-0.73$, we find that $p_l$ decreases as a stretched exponential with the distance between sites. In contrast, it decreases as a power law in the non-interacting case. Thus, we conclude that the interaction case is quasilocal in the asysmptotic disorder strength. The other extreme of relevance is $\vert \Delta \vert \gg 1 $ which will have $p_{l\neq 0} =0$. 

\section{Sub-Poissonian level statistics \label{sec:lev}}
The sub-Poissonian nature of the mean gap ratio can be understood within the RSRG-X procedure in the following way: in the strong disorder regime for finite spins, the energy gaps of the local decimated operators are sparsely distributed within a broad range, thus leading to clustering of eigenvalues, as shown in Fig. 2(c,d) in the letter. In addition, the spin-frozen domains could lead to quasi-degenerate eigenvalues that differ in energy with higher order perturbation; part of the degeneracy is taken care of by choice of the spin inversion symmetry sector but could still be present if there are multiple domains of frozen spins which preserve total magnetization on inverting spins of different domains.

We directly look into the unfolded adjacent energy level gap to quantify the sub-Poissonian behavior. We use a non-standard unfolding procedure to take into consideration the presence of clustered eigenvalues in the DOS at a strong disorder limit.
The unfolding is done by fitting with a positive cubic spline on $M$ equidistant points between $E_{\min}$ and $E_{\max}$, giving a function $f(E)$. The unfolded eigenvalues, given as $e_i=f(E_i)$, define the energy level gaps $\delta_i \equiv e_{i+1}-e_i$, with mean level spacing unity $\braket{\delta_i}=1$.
We consider all eigenvalues for system sizes $M \in [10,12,14,16]$ with the number of disorder realizations $[2\cdot 10^5,5\cdot 10^4,10^4,4\cdot 10^3]$. When making comparisons of quantities 
for different $M$,
to avoid energy levels from the edges of the spectrum, we use a Gaussian filter $\rho(e_i)=\exp\left(-\frac{(e_i-\bar{e})^2}{2(\mu\sigma)^2} \right)$, where $\bar{e}$ is the average unfolded energy and $\sigma$ is the standard deviation with $\eta = 0.3$. 

\begin{figure}[!h]
\begin{center}
\includegraphics[width=\linewidth]{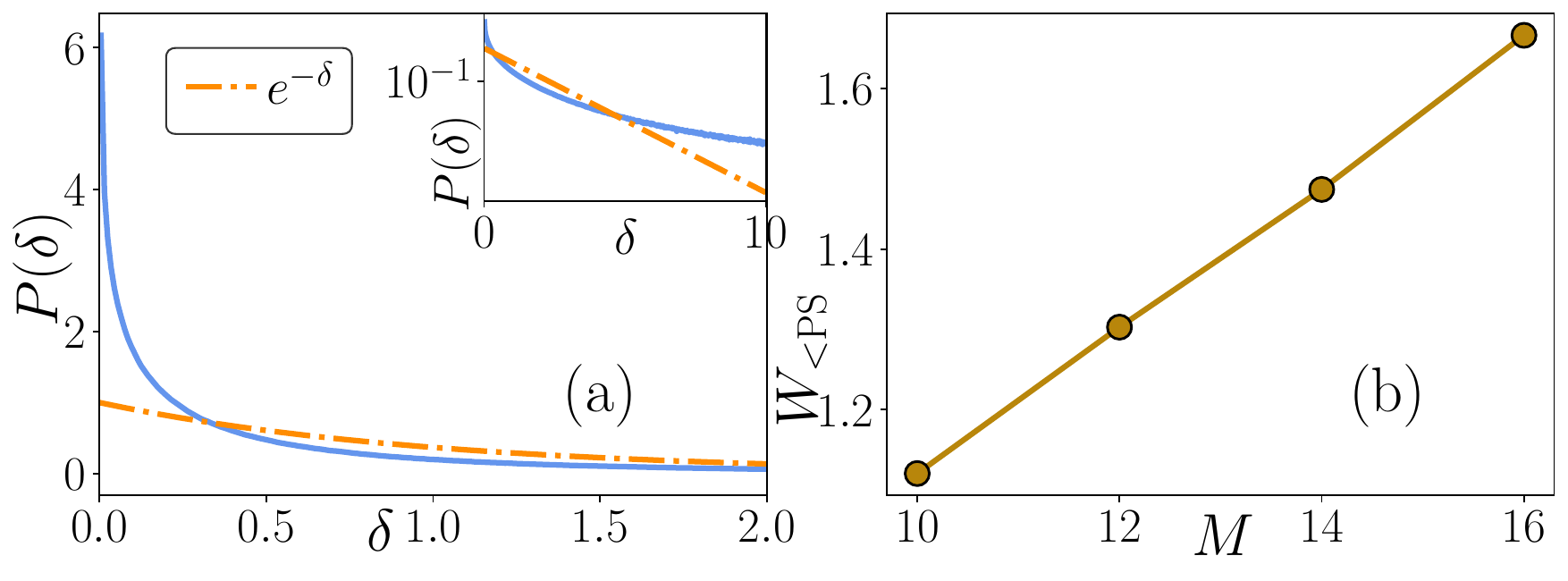}\\
\caption{(a) Probability density function for energy gaps with $M=16$ and $W=2$. The dashed-dotted orange curve gives the Poisson curve $e^{-\delta}$. The inset here is the same plot with the log scale on the y-axis. (b) The drift with disorder strength for a sub-Poissonian value of mean gap ratio ($\bar{r}\approx 0.34$), given as $W_{<PS}$, with increasing $M$.}\label{fig:lev}
\end{center}
\end{figure}

The quasi-degenerate eigenvalues 
yield exceptionally small energy level gaps, while large level gaps occur for adjacent energy levels in different eigenvalue clusters. In Fig. \ref{fig:lev}(a) and its inset, both scenarios are compared with the Poisson distribution of energy gaps. With an increasing number of spins, the local energy gaps are more evenly distributed within a broad range, and we would expect a weaker sub-Poissonian nature. For a small number of spins, the disorder strength for a given sub-Poissonian value ($\bar{r} \approx 0.34 $), $W_{<PS}$, increases linearly with system size, as shown in Fig. \ref{fig:lev}(b), we do not extrapolate our claims to the thermodynamic limit from the very small $M$ values we consider.

To quantify the level attraction of energy gaps in the strong disorder regime, we consider the cumulative probability density function for the minimum energy gap, $\mathrm{C}[\delta_{\mathrm{min}}]$, motivated by \cite{Imbrie16} minimum level attraction condition, i.e., the probability that the minimum gap for a given disorder realization $\delta_{\min}\equiv \min\lbrace \delta_i\rbrace$ is below a particular value $\delta$ is less than a power law condition given as \cite{Morningstar22}
\begin{equation}
    \mathrm{C}[\delta]\equiv P[\delta_{\min}<\delta]<\delta^{\beta + 1}A_0^L.
\end{equation}
Using the parametrization $\mathrm{C}[\delta]=\delta^{\beta}A_0^L$ we extract the $\beta$ values. Until we reach the Poissonian from the GOE limit, the constant $A_0^L$ is approximated by the Hilbert space dimension, $\mathcal{N}$, while for the sub-Poissonian regime (we consider here $W>0.7$), $A_0^L$ decreases with increasing disorder strength.

\begin{figure}[!h]
\begin{center}
\includegraphics[width=\linewidth]{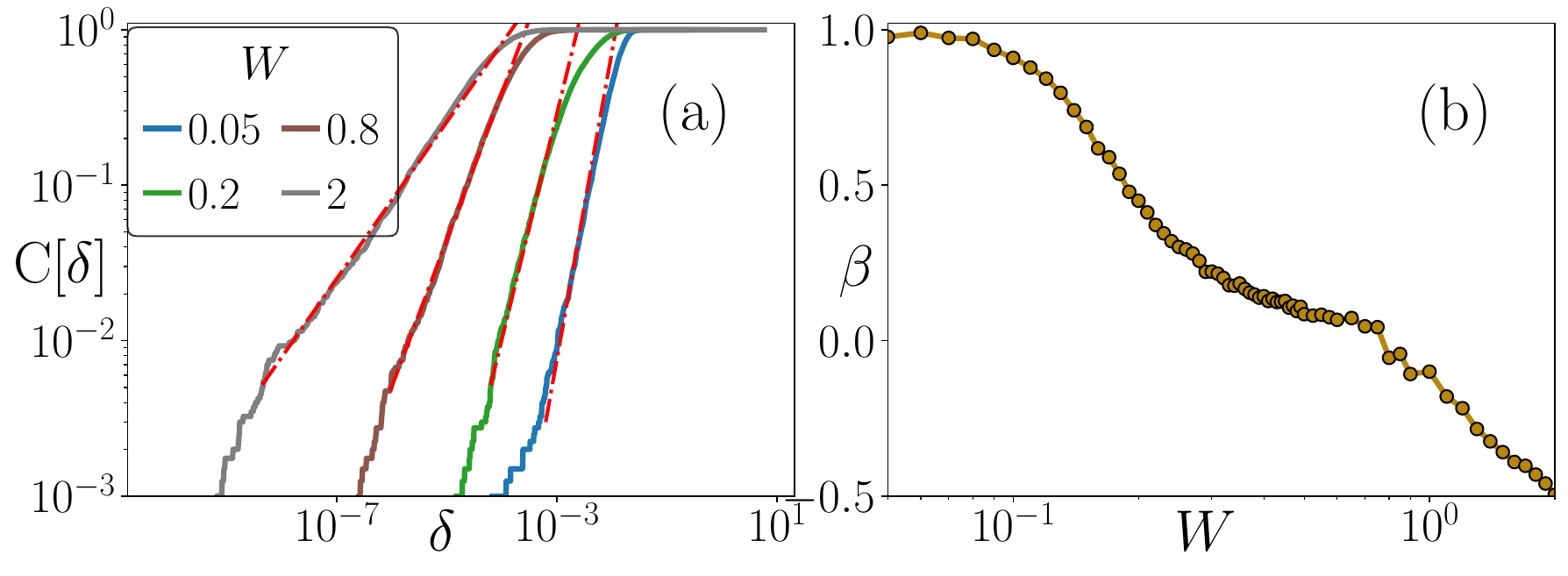}\\
\caption{(a) Cumulative probability density function for a minimum of energy gap $\mathrm{C}[\delta]$ with respect to energy gaps for different disorder strengths $W$. The red dashed-dotted curve shows a fit with a power law function. (b) The exponents $\beta$ extracted from fitting $\mathrm{C}[\delta]$ to a power law distribution $\propto \delta^{\beta+1}$. Both (a) and (b) consider $M=16$ and $10^3$ disorder realizations.}\label{fig:levmin}
\end{center}
\end{figure}

In Fig. \ref{fig:levmin}(a), we show the fits of $\mathrm{C}[\delta]$ for $4$ different disorder strengths for $M=16$. In Fig. \ref{fig:levmin}(b), we show the extracted exponents $\beta$ varying with disorder strength. There is a monotonous decrease in $\beta$ from the Dyson index for GOE on increasing disorder strength from $\beta \approx 1$ in the ergodic regime to  $\beta \approx 0$ for the Poissonian and $\beta < 0$ for the sub-Poissonian regime.

\section{Spectral form factor\label{sec:sff}}
We briefly describe the nature of the spectral form factor found for the random bond XXZ model. The spectral form factor (SFF) is defined as a Fourier transform of spectral two-point correlation function 
\begin{equation}
    \mathcal{K}(\tau)=\frac{1}{\eta}\left\langle{\left\vert \sum_{i=1}^{\mathcal{N}}\rho(e_i)e^{-i2\pi e_i \tau} \right\vert^2}\right\rangle
\end{equation}
where $e_i$ are the unfolded eigenvalues, $\rho$ is the smooth filter to avoid the influence of spectral edges, $\eta$ is the Hilbert space dimension, and $\braket{}$ is the disorder averaging with normalization factor $\eta=\braket{\sum_i \vert \rho(e_i)\vert^2 }$. To avoid energy levels from the edges of the spectrum we use a gaussian filter $\rho(e_i)=\exp\left(-\frac{(e_i-\bar{e})^2}{2(\mu\sigma)^2} \right)$, where $\bar{e}$ is the average unfolded energy and $\sigma$ is the standard deviation with $\eta = 0.3$. The mean level spacing  is unity $\braket{\delta_k}$ once unfolded making the Heisenberg time, $t_H=1/\braket{\delta_k}$, unityindependent of disorder strength $\tau_H =1$.

For a Gaussian orthogonal ensemble (GOE), the SFF, $\mathcal{K}(\tau)$, is given by RMT as 
\begin{equation}
\mathcal{K}_{\mathrm{GOE}}(\tau)=\begin{cases}
    2\tau -\tau \ln(1+2\tau), & \text{if $\tau\leq 1$},\\
    2-\tau \ln (\frac{2\tau+1}{2\tau-1}), & \text{if $\tau >1$}.
  \end{cases}
\end{equation}
The Thouless time, $t_{Th}$, is 
defined as the time at which the spectral form factor approaches the spectral form factor of GOE.
Here, we give the Thouless time in the scale of Heisenberg time $\tau_{Th} \equiv t_{Th}/t_{H} $ 
as the time in which 
\begin{equation}
    \delta \mathcal{K}(\tau)=\left\vert \log\left(\frac{\mathcal{K}(\tau)}{\mathcal{K}_{\mathrm{GOE}}(\tau)}\right) \right \vert 
\end{equation}
$\delta \mathcal{K}(\tau) $ approaches a small positive cutoff $\delta \mathcal{K}(\tau) \approx 0.1$.

\begin{figure}[!h]
\begin{center}
\includegraphics[width=\linewidth]{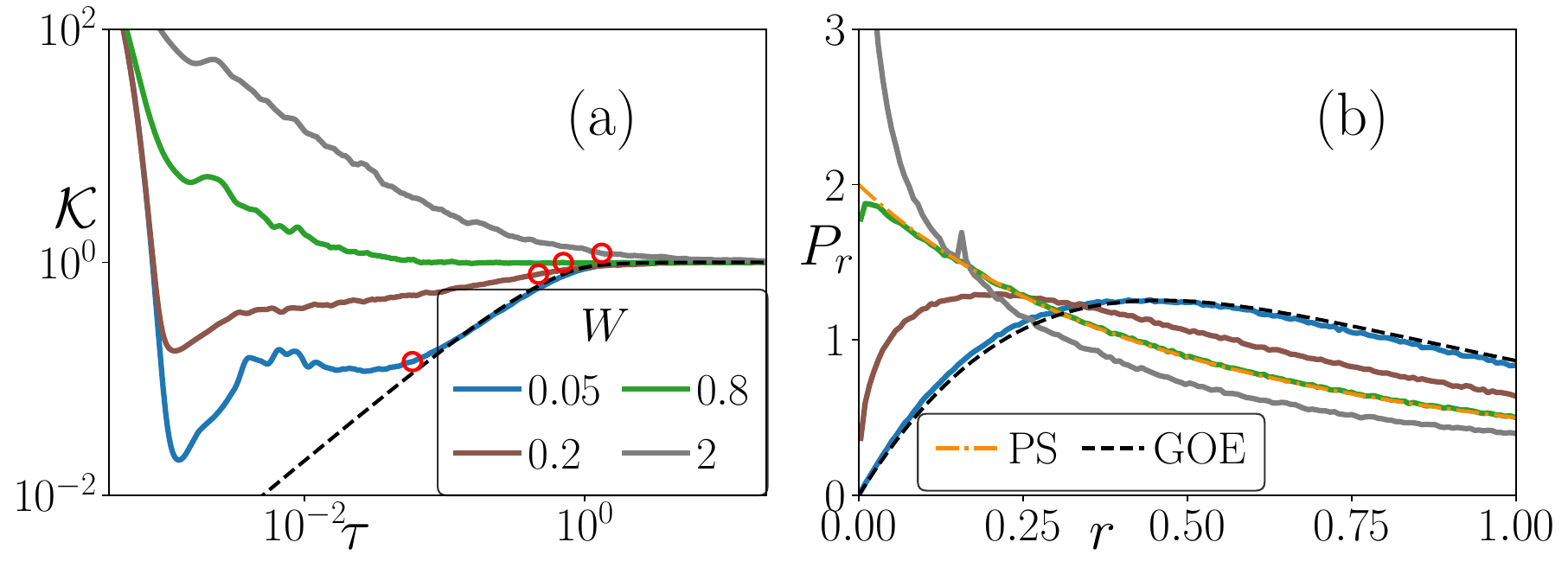}\\
\caption{(a) Spectral form factor ($\mathcal{K}$) with time scaled in units of Heisenberg time ($\tau\equiv t/t_{H}$) for disorder strengths representing different regimes, in comparison with the GOE value shown as the dashed curve. The red circles denote the points where $\delta \mathcal{K}(\tau)\approx0.1$. (b) The distribution of mean gap ratio ($P_r$) for the disorder strengths considered in (a), the dashed-dotted curve is the curve from the Poisson distribution. Both (a) and (b) consider $M=16$ and $10^3$ disorder realizations.}\label{fig:SFF_dif_reg}
\end{center}
\end{figure}
For the random bond XXZ model \eqref{eq::Ham}, we explore the SFF characteristics in different regimes, namely in the ergodic regime (we take $W=0.05$), which RMT approximates, a midpoint between ergodic and MBL regime ($W=0.2$), the MBL regime where the mean gap ratio is Poisson distributed ($W=0.8$) and finally a sub-Poissonian value for mean gap ratio ($W=2$) in Fig. \ref{fig:SFF_dif_reg}. 
In the ergodic regime, after initial small times dependent on the particular form of the Hamiltonian, approach the GOE curve of $\mathcal{K}(\tau)$ following the increasing ramp behavior. 
For $W=0.2$, the  GOE curve is approached at later times. In the MBL regime, where the eigenvalues are uncorrelated and the energy gaps are Poisson distributed, the curve is flat as it approaches $\mathcal{K}(\tau)=1$ \cite{Prakash21}. Finally, for the sub-Poissonian regime, the presence of level attraction leads to a decreasing SFF with time with values $\mathcal{K}(\tau)>1$ until it reaches the $\mathcal{K}_{\mathrm{GOE}}(\tau)$ curve.

\begin{figure}[!h]
\begin{center}
\includegraphics[width=\linewidth]{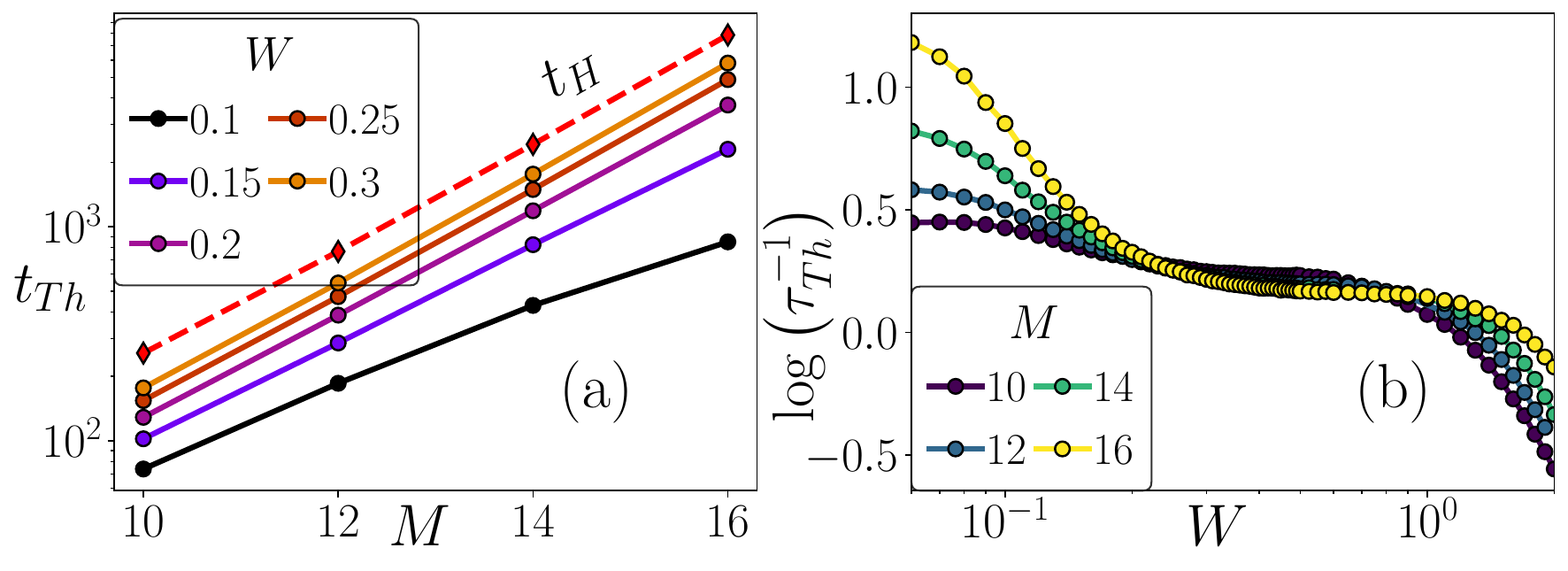}\\
\caption{(a) Thouless time ($t_{Th}$) varying with $M$ for different disorder strengths. The diamond-shaped dashed red curve represents the Heisenberg time, $t_H$, for disorder strength $W=0.2$. (b) The logarithm of inverse Thouless time in Heisenberg time scale $\log{(\tau_{Th}^{-1}})$ with respect to $W$ for different $M$.}\label{fig:Thouless}
\end{center}
\end{figure}

For the 3-D Anderson model, the Thouless time has weaker dependence with $M$ than the Heisenberg time \cite{Sierant20b,Suntajs21} for small disorder strengths, which eventually reaches similar scaling with increasing disorder strength. This behavior is also seen in XXZ spin-1/2 chain \cite{Sierant20p}. In Fig. \ref{fig:Thouless}(a), we find the Heisenberg time scales exponentially with $M$ for small disorder strengths while the Thouless time scales in a weaker fashion, which eventually scales with the same scaling at an intermediate disorder strength. 

In 3-D Anderson and Quantum Sun models \cite{Suntajs21,Suntajs22,Pawlik23}, the crossing points of disorder strength were accurately predicted with the condition 
\begin{equation}
    \frac{t_H}{t_{Th}} = \mathrm{const.}
\end{equation}
Exploring the above condition with $\log{(\tau_{Th}^{-1}})$ we find characteristically similar behavior to mean gap ratio, $\bar{r}$, the crossing point around $W\approx 0.25$, and regime that corresponds with sub-Poissonian $\bar{r}$,  $W>1$ where Thouless time is greater than the Heisenberg time.

\subsection{ Description of numerics\label{sec:num}}
\subsubsection{Diagonalization}
Statistical observables, namely the mean gap ratio, $\bar{r}$, and the rescaled average half-chain EE, $S$, required the extraction of eigenenergies and eigenstates of the Hamiltonian in Eq. \ref{eq::pd_Ham_lr} at high energies ($\epsilon = 0.333$). Note that while extracting high energy states in the main letter as in Fig. 2(a) and Fig. 4(b) we consider again states around $\epsilon = 0.333$ with half the disorder realization as in Table \ref{table:disord}.  For a system with a small number of spins $M\leq16$, we use the standard exact diagonalization (ED) method for dense matrices, while for $M\geq18$ the
recently developed state-of-the-art method for sparse matrices \textit{polynomially filtered exact diagonalization}
(POLFED) \cite{Sierant20p} was used.

\subsection{Time evolution}
For the time evolution of a system with a small number of spins, i.e., $M\leq 16$, we employed exact diagonalization to make a time evolution. In contrast, for larger particle numbers, we employ the matrix product state (MPS) based time-dependent variational principle (TDVP) method \cite{Haegeman11,Koffel12,Haegeman16}. Specifically, we use the `hybrid' scheme of TDVP \cite{Paeckel19,Goto19,Chanda20t,Chanda20m} with step size $\delta t  = 0.1$, where the two-site variant of TDVP is initially used to dynamically grow the bond dimension of the MPS until the dimension of the largest MPS bond reaches a predetermined value $\chi_{max}$. For subsequent dynamics, a one-site variant of TDVP is employed. In our calculations, we set the maximum value the bond dimension can grow to as $\chi_{max}=1024$, while the bond dimensions do not reach the value $\chi_{max}$  for the maximum time up to which we make the time evolution $t = 5000$. 

\subsubsection{Parameters for cost-function minimization}
To minimize the cost function, we employ the differential evolution method implemented in SciPy \cite{Scipy}. We run a series of independent differential evolution algorithms with different random seeds for each $\mathcal{C}_X$ shown in \ref{table:CF}. We employ $1000$ independent realizations to find the optimal minimization. In each realization, we allow up to $10^4$ iterations with a relative tolerance of convergence $10^{-4}$ and a population size of $250$.
\subsubsection{Calculation of errors}
\begin{figure}[!h]
\begin{center}
\includegraphics[width=.9\linewidth]{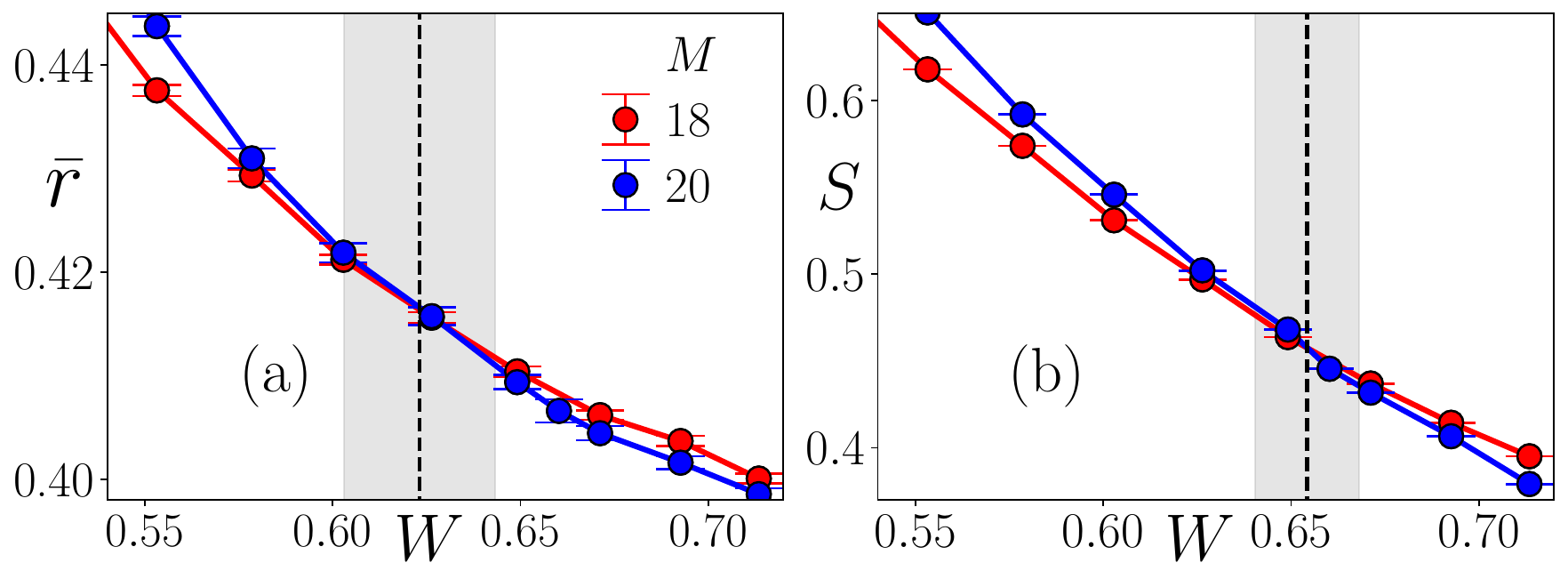}\\
\caption{An example of extracting the crossing point $W_{\times}$ between systems with two different numbers of spins $M$ for both (a) the mean gap ratio $\bar{r}$ and (b) the average rescaled half-chain EE $S$. The black dashed line denotes the identified crossing point, while the grayed part signifies the error margins.}\label{fig:Cross_cal}
\end{center}
\end{figure}
After averaging over different disorder realizations, the statistical errors associated with the mean value of observables $\bar{X}$ are calculated by initially averaging over the disorder realization, $X_s$, then identifying the uncertainty \cite{Sierant23}  
\begin{equation}
    \sigma_X=\frac{(\langle (X_s-\bar{X})^2\rangle )^{1/2}}{N_{dis}^{1/2}}
\end{equation}
here $N_{dis}$ is the number of disorder realizations.

To calculate the errors associated with the disorder strengths $W_T^X(M)$ and  $W_{\times}^X(M)$ and the errors associated with fitting them to a functional form, we consider the random sampling of errors that are assumed to be Gaussian distributed. For example, to calculate the crossing point disorder strength   $W_{\times}^X(M)$ and the associated errors. Resample from a Gaussian distribution with the standard deviation given as $\sigma_X$ around a mean value $\bar{X}$ and fit the associated curves of the samples of system sizes $M-1$ and $M+1$ in the transition regime with a $5$th order polynomial and extract the point where they cross. Re-sampling over a $10000$ times, we find a distribution of intercepts, the mean value of which gives the associated crossing point, while the standard deviation gives the associated error as shown in Fig. \ref{fig:Cross_cal}. A similar procedure calculates the error associated with $W_T^X(M)$ with a cubic spline instead of a polynomial fit. While the errors associated with the fits, shown as the gray area in \ref{fig:W_cr}, are extracted by resampling and fitting with the functional form $W_{\times}^X(M)=W_{\infty}+a_1/M+a_2/M^2$.

\vfill\eject
\end{document}